\newif{\ifchangetext}
\changetextfalse

\documentclass[iop]{emulateapj}
\slugcomment{}
\journalinfo{accepted to ApJ}  

\usepackage{ifthen}
\usepackage{natbib}   
\usepackage{amsmath}  
\usepackage{hyperref}      
\usepackage[usenames,dvipsnames]{xcolor}  

\ifchangetext
  \newcommand{\change}[1]{{ \textcolor{blue}{#1}}}
  \newcommand{\changenote}[1]{{ \textcolor{blue}{#1}}}

\else
  \newcommand{\change}[1]{#1}

  \newcommand{\changenote}[1]{}
\fi

\DeclareGraphicsExtensions{.pdf,.png,.jpg}

\defcitealias{Patel:2014}{P14}
\def\P14{\citetalias{Patel:2014}}

\def\tomas{HFF14Tom}

\def\Om{\ensuremath{\Omega_{\rm m}}}

\def\OL{\ensuremath{\Omega_{\Lambda}}}

\def\LCDM{$\Lambda$CDM}

\def\Ho{\ensuremath{H_0}}

\def\arcsec{\ensuremath{^{\prime\prime}}}

\def\Msun{\mbox{M$_{\odot}$}}

\newcommand{\CCSN}{core collapse SN}
\newcommand{\CCSNe}{core collapse SNe}
\newcommand{\SNIa}{Type Ia SN}
\newcommand{\SNeIa}{Type Ia SNe}

\def\HST{{\it HST}}

\shorttitle{A Type Ia SN Behind Abell 2744}
\shortauthors{Rodney et al.}

\begin{document}

\title{Illuminating a Dark Lens : A Type Ia Supernova Magnified by \\ the Frontier Fields Galaxy Cluster Abell 2744}

\newcommand{\USC}{Department of Physics and Astronomy, University of South Carolina, 712 Main St., Columbia, SC 29208, USA}
\newcommand{\HubbleFellow}{Hubble Fellow}
\newcommand{\JHU}{Department of Physics and Astronomy, The Johns Hopkins University, 3400 N. Charles St., Baltimore, MD 21218, USA}
\newcommand{\STScI}{Space Telescope Science Institute, 3700 San Martin Dr., Baltimore, MD 21218, USA}
\newcommand{\Chicago}{Department of Physics, The University of Chicago, Chicago, IL 60637, USA}
\newcommand{\Kavli}{Kavli Institute for Cosmological Physics, University of Chicago, 5640 South Ellis Avenue Chicago, IL 60637}
\newcommand{\Berkeley}{Department of Astronomy, University of California, Berkeley, CA 94720-3411, USA}
\newcommand{\LBNL}{Lawrence Berkeley National Laboratory, Berkeley, CA 94720, USA}
\newcommand{\Riverside}{Department of Physics and Astronomy, University of California, Riverside, CA 92521, USA}
\newcommand{\Andalucia}{Instituto de Astrof\'isica de Andaluc\'ia (CSIC), E-18080 Granada, Spain}
\newcommand{\Cantabria}{IFCA, Instituto de F\'isica de Cantabria (UC-CSIC), Av. de Los Castros s/n, 39005 Santander, Spain}
\newcommand{\IFCA}{\Cantabria}
\newcommand{\SaoPaulo}{Instituto de Astronomia, Geof\'isica e Ci\^encias Atmosf\'ericas, Universidade de S\~ao Paulo, Cidade Universit\'aria, 05508-090, S\~ao Paulo, Brazil}
\newcommand{\WKU}{Department of Physics, Western Kentucky University, Bowling Green, KY 42101, USA}
\newcommand{\AMNH}{Department of Astrophysics, American Museum of Natural History, Central Park West and 79th Street, New York, NY 10024, USA}
\newcommand{\NYU}{Center for Cosmology and Particle Physics, New York University, New York, NY 10003, USA}
\newcommand{\DARK}{Dark Cosmology Centre, Niels Bohr Institute, University of Copenhagen, Juliane Maries Vej 30, DK-2100 Copenhagen, Denmark}
\newcommand{\Copenhagen}{\DARK}
\newcommand{\Arizona}{Department of Astronomy, University of Arizona, Tucson, AZ 85721, USA}
\newcommand{\SantaCruz}{Department of Astronomy and Astrophysics, University of California, Santa Cruz, CA 92064, USA}
\newcommand{\NotreDame}{Department of Physics, University of Notre Dame, Notre Dame, IN 46556, USA}
\newcommand{\TelAviv}{Department of Astrophysics, Tel Aviv University, 69978 Tel Aviv, Israel}
\newcommand{\Rutgers}{Department of Physics and Astronomy, Rutgers, The State University of New Jersey, Piscataway, NJ 08854, USA}
\newcommand{\CfA}{Harvard/Smithsonian Center for Astrophysics, Cambridge, MA 02138, USA}
\newcommand{\Minnesota}{School of Physics and Astronomy, University of Minnesota, 116 Church Street SE, Minneapolis, MN 55455, USA}
\newcommand{\NOAO}{National Optical Astronomical Observatory, Tucson, AZ 85719, USA}
\newcommand{\UCSB}{Department of Physics, University of California, Santa Barbara, CA 93106-9530, USA}
\newcommand{\SantaBarbara}{\UCSB}
\newcommand{\LCOGT}{Las Cumbres Observatory Global Telescope Network, 6740 Cortona Dr., Suite 102, Goleta, California 93117, USA}
\newcommand{\Colby}{Colby College, 4000 Mayflower Hill Dr, Waterville, ME 04901, USA}
\newcommand{\Kentucky}{University of Kentucky, Lexington, KY 40506}
\newcommand{\UCDavis}{University of California Davis, 1 Shields Avenue, Davis, CA 95616}
\newcommand{\UCLA}{Department of Physics and Astronomy, University of California, Los Angeles, CA 90095}
\newcommand{\Packard}{Packard Fellow}
\newcommand{\CalTech}{Cahill Center for Astronomy and Astrophysics, California Institute of Technology, MC 249-17, Pasadena, CA 91125, USA}
\newcommand{\Durham}{Institute for Computational Cosmology, Durham University, South Road, Durham DH1 3LE, UK}
\newcommand{\KwaZulu}{Astrophysics and Cosmology Research Unit, School of Mathematical Sciences, University of KwaZulu-Natal, Durban 4041, South Africa}
\newcommand{\HongKong}{Department of Physics, The University of Hong Kong, Pokfulam Road, Hong Kong}
\newcommand{\Oxford}{Department of Physics, University of Oxford, Keble Road, Oxford OX1 3RH, UK}
\newcommand{\CRAL}{CRAL, Observatoire de Lyon, Universit\'e Lyon 1, 9 Avenue Ch. Andr\'e, F-69561 Saint Genis Laval Cedex, France}
\newcommand{\Michigan}{Department of Astronomy, University of Michigan, 1085 S. University Avenue, Ann Arbor, MI 48109, USA}
\newcommand{\Hebrew}{The Hebrew University, The Edmond J. Safra Campus - Givat Ram, Jerusalem 9190401, Israel}
\newcommand{\IllinoisAstro}{ Astronomy Department, University of Illinois at Urbana-Champaign, 1002 W.\ Green Street, Urbana, IL 61801, USA }
\newcommand{\IllinoisPhysics}{ Department of Physics, University of Illinois at Urbana-Champaign, 1110 W.\ Green Street, Urbana, IL 61801, USA }
\newcommand{\INAF}{INAF, Osservatorio Astronomico di Bologna, via Ranzani 1, I-40127 Bologna, Italy}
\newcommand{\JPL}{Jet Propulsion Laboratory, California Institute of Technology, 4800 Oak Grove Drive, Pasadena, CA 91109, USA}
\newcommand{\INFN}{INFN, Sezione di Bologna, Viale Berti Pichat 6/2, I-40127 Bologna, Italy}
\newcommand{\EHU}{Fisika Teorikoa, Zientzia eta Teknologia Fakultatea, Euskal Herriko Unibertsitatea UPV/EHU}
\newcommand{\Basque}{IKERBASQUE, Basque Foundation for Science, Alameda Urquijo, 36-5 48008 Bilbao, Spain}

\newcounter{affilct}
\setcounter{affilct}{0}

\makeatletter
\newcommand{\affilref}[1]{%
  \@ifundefined{c@#1}%
    {\newcounter{#1}%
     \setcounter{#1}{\theaffilct}%
     \refstepcounter{affilct}%
     \label{#1}%
     }{}%
  \ref{#1}%
 }
\makeatother

\makeatletter
\newcommand*\affilreftxt[2]{%
  \@ifundefined{c@#1txt}
    {\newcounter{#1txt}%
     \setcounter{#1txt}{1}
     \altaffiltext{\ref{#1}}{#2}
     }{
     }
  }
\makeatother

\author{Steven~A.~Rodney\altaffilmark{\affilref{USC},\affilref{JHU},\affilref{HubbleFellow}}}
\affilreftxt{USC}{\USC}
\affilreftxt{JHU}{\JHU}
\affilreftxt{HubbleFellow}{\HubbleFellow}
\email{srodney@sc.edu}

\author{Brandon~Patel\altaffilmark{\affilref{Rutgers}}}
\affilreftxt{Rutgers}{\Rutgers}

\author{Daniel~Scolnic\altaffilmark{\affilref{Kavli}}}
\affilreftxt{Kavli}{\Kavli}

\author{Ryan~J.~Foley\altaffilmark{\affilref{IllinoisPhysics},\affilref{IllinoisAstro}}}
\affilreftxt{IllinoisPhysics}{\IllinoisPhysics}
\affilreftxt{IllinoisAstro}{\IllinoisAstro}

\author{Alberto Molino\altaffilmark{\affilref{Andalucia},\affilref{SaoPaulo}}}
\affilreftxt{Andalucia}{\Andalucia}
\affilreftxt{SaoPaulo}{\SaoPaulo}

\author{Gabriel Brammer\altaffilmark{\affilref{STScI}}}
\affilreftxt{STScI}{\STScI}

\author{Mathilde Jauzac\altaffilmark{\affilref{Durham},\affilref{KwaZulu}}}
\affilreftxt{Durham}{\Durham}
\affilreftxt{KwaZulu}{\KwaZulu}

\author{Maru\v{s}a Brada\v{c}\altaffilmark{\affilref{UCDavis}}}
\affilreftxt{UCDavis}{\UCDavis}

\author{Tom Broadhurst\altaffilmark{\affilref{EHU},\affilref{Basque}}}
\affilreftxt{EHU}{\EHU}
\affilreftxt{Basque}{\Basque}

\author{Dan Coe\altaffilmark{\affilref{STScI}}}
\affilreftxt{STScI}{\STScI}

\author{Jose~M.~Diego\altaffilmark{\affilref{Cantabria}}}
\affilreftxt{Cantabria}{\Cantabria}

\author{Or Graur\altaffilmark{\affilref{NYU},\affilref{AMNH}}}
\affilreftxt{NYU}{\NYU}
\affilreftxt{AMNH}{\AMNH}

\author{Jens Hjorth\altaffilmark{\affilref{DARK}}}
\affilreftxt{DARK}{\DARK}

\author{Austin Hoag\altaffilmark{\affilref{UCDavis}}}
\affilreftxt{UCDavis}{\UCDavis}

\author{Saurabh W.~Jha\altaffilmark{\affilref{Rutgers}}}
\affilreftxt{Rutgers}{\Rutgers}

\author{Traci L.~Johnson\altaffilmark{\affilref{Michigan}}}
\affilreftxt{Michigan}{\Michigan}

\author{Patrick Kelly\altaffilmark{\affilref{Berkeley}}}
\affilreftxt{Berkeley}{\Berkeley}

\author{Daniel Lam\altaffilmark{\affilref{HongKong}}}
\affilreftxt{HongKong}{\HongKong}

\author{Curtis McCully\altaffilmark{\affilref{LCOGT},\affilref{UCSB}}}
\affilreftxt{LCOGT}{\LCOGT}
\affilreftxt{UCSB}{\UCSB}

\author{Elinor Medezinski\altaffilmark{\affilref{JHU},\affilref{Hebrew}}}
\affilreftxt{JHU}{\JHU}
\affilreftxt{Hebrew}{\Hebrew}

\author{Massimo Meneghetti\altaffilmark{\affilref{INAF},\affilref{JPL},\affilref{INFN}}}
\affilreftxt{INAF}{\INAF}
\affilreftxt{JPL}{\JPL}
\affilreftxt{INFN}{\INFN}

\author{Julian~Merten\altaffilmark{\affilref{Oxford}}}
\affilreftxt{Oxford}{\Oxford}

\author{Johan Richard\altaffilmark{\affilref{CRAL}}}
\affilreftxt{CRAL}{\CRAL}

\author{Adam Riess\altaffilmark{\affilref{JHU},\affilref{STScI}}}
\affilreftxt{JHU}{\JHU}
\affilreftxt{STScI}{\STScI}

\author{Keren Sharon\altaffilmark{\affilref{Michigan}}}
\affilreftxt{Michigan}{\Michigan}

\author{Louis-Gregory Strolger\altaffilmark{\affilref{STScI},\affilref{WKU}}}
\affilreftxt{STScI}{\STScI}
\affilreftxt{WKU}{\WKU}

\author{Tommaso Treu\altaffilmark{\affilref{UCLA},\affilref{Packard}}}
\affilreftxt{UCLA}{\UCLA}
\affilreftxt{Packard}{\Packard}

\author{Xin Wang\altaffilmark{\affilref{UCSB}}}
\affilreftxt{UCSB}{\UCSB}

\author{Liliya~L.~R.~Williams\altaffilmark{\affilref{Minnesota}}}
\affilreftxt{Minnesota}{\Minnesota}

\author{Adi Zitrin\altaffilmark{\affilref{CalTech},\affilref{HubbleFellow}}}
\affilreftxt{CalTech}{\CalTech}

\begin{abstract}
{
SN \tomas\ is a Type Ia Supernova (SN) discovered at
$z=1.3457\pm0.0001$ behind the galaxy cluster Abell 2744 ($z=0.308$).
In a cosmology-independent analysis, we find that \tomas\ is
$0.77\pm0.15$ magnitudes brighter than unlensed Type Ia SNe at similar
redshift, implying a lensing magnification of $\mu_{\rm
obs}=2.03\pm0.29$.  This observed magnification provides a rare
opportunity for a direct empirical test of galaxy cluster lens models.
Here we test 17 lens models, 13 of which were generated before the SN
magnification was known, qualifying as pure ``blind tests''.  The
models are collectively fairly accurate: 8 of the models deliver
median magnifications that are consistent with the measured $\mu$ to
within 1$\sigma$.  However, there is a subtle systematic bias: the
significant disagreements all involve models {\it overpredicting} the
magnification. We evaluate possible causes for this mild bias, and
find no single physical or methodological explanation to account for
it.  We do find that model accuracy can be improved to some extent
with stringent quality cuts on multiply-imaged systems, such as
requiring that a large fraction have spectroscopic redshifts.  In
addition to testing model accuracies as we have done here, Type Ia SN
magnifications could also be used as inputs for future lens models of
Abell 2744 and other clusters, providing valuable constraints in
regions where traditional strong- and weak-lensing information is
unavailable.}
\end{abstract}


\section{Introduction}
\label{sec:Introduction}

Galaxy clusters can be used as cosmic telescopes to magnify distant
background objects through gravitational lensing, which can
substantially increase the reach of deep imaging surveys.  The lensing
magnification enables the study of objects that would otherwise be
unobservable because they are either intrinsically
faint \citep[e.g.][]{Schenker:2012,Alavi:2014} or extremely
distant \citep[e.g.][]{Franx:1997,Ellis:2001,Hu:2002,Kneib:2004,Richard:2006,Richard:2008,Bouwens:2009a,Maizy:2010,Zheng:2012,Coe:2013,Bouwens:2014,Zitrin:2014b}.
Background galaxies are also {\it spatially} magnified, allowing for
studies of the internal structure of galaxies in the early universe
with resolutions of $\sim$100
pc \citep[e.g.][]{Stark:2008,Jones:2010,Yuan:2011,Wuyts:2014,Livermore:2015}.

Gravitational lensing can also provide a powerful window onto the {\it
transient} sky through an appropriately cadenced imaging survey.  The
flux magnification from strong-lensing clusters is especially valuable
for the study of $z>1.5$ supernovae
(SNe) \citep[e.g.][]{Kovner:1988,Kolatt:1998,Sullivan:2000,Saini:2000,Gunnarsson:2003,Goobar:2009,Postman:2012},
which are still extremely difficult to characterize in unlensed
fields \citep[e.g.][]{Riess:2001,Riess:2007,Suzuki:2012,Rodney:2012,Rubin:2013,Jones:2013}.

In the case of lensed SNe, we can also reverse the experimental setup:
instead of using strong-lensing clusters to study distant SNe, we can
use the SNe as tools for examining the lenses \citep{Riehm:2011}.  The
most valuable transients for testing and improving cluster lens models
would be {\it strongly lensed} SNe that are resolved into multiple
images\ \citep{Holz:2001,Oguri:2003}.  Transients that are lensed into
multiple images can also become cosmological tools, as the measurement
of time delays between the images can provide cosmographic information
to constrain the Hubble parameter \citep{Refsdal:1964} and other
cosmological parameters \citep{Linder:2011}.  We have recently
observed the first example of a multiply-imaged SN \citep{Kelly:2015}.
We expect to detect the reappearance of this object (called ``SN
Refsdal'') within the next
year \citep{Oguri:2015,Sharon:2015,Diego:2015}, delivering a precise
test of lens model predictions.  Although detections of such objects
are currently very unlikely \citep{Li:2012}, they will become much
more common in the next decade \citep{Coe:2009,Dobke:2009}, and may be
developed into an important new cosmological
tool \citep{Oguri:2010,Linder:2011}.

In addition to time delays from multiply-imaged SNe, we can also
put cluster mass models to the test with the much more common category
of Type Ia SNe that are magnified but not multiply-imaged.  
\citet[][hereafter P14]{Patel:2014} and \citet{Nordin:2014} presented
independent analyses of three lensed SNe, of which at least 2 are
securely classified as Type Ia SNe -- all found in the Cluster Lensing
and Supernova survey with Hubble (CLASH, PI:Postman, HST Program ID
12068, \citealt{Postman:2012}).  Both groups demonstrated that these
standard candles can be used to provide accurate and precise
measurements of the true absolute magnification along a random sight
line through the cluster.  Although in these cases the SNe were used
to {\it test} the cluster mass models, one could in principle
incorporate the measured magnifications of Type Ia SNe into the
cluster as additional model constraints.  In that role, Type Ia SNe
have the particular value that they can be found anywhere in the
cluster field.  Thus, they can deliver model constraints in regions of
``middle distance'' from the cluster core, where both strong- and
weak-lensing constraints are unavailable. Moreover, given
enough time, multiple background \SNeIa\ could be measured behind the
same cluster, each providing a new magnification constraint.

One of the key values in observing standard candles behind
gravitational lenses is in addressing the problem of the mass-sheet
degeneracy \citep{Falco:1985,Schneider:1995}.  This degeneracy arises
because one can introduce into a lens model an unassociated sheet of
uniform mass in front of or behind the lens, without disturbing the
primary observable quantities.  For example, take a lens model with a
given surface mass density $\kappa$, and then transform the surface
mass density to $\kappa^{\prime}=(1-\lambda)\kappa+\lambda$ for any
arbitrary value $\lambda$.  Both the $\kappa$ and $\kappa^{\prime}$
models will produce exactly the same values for all positional and
shear constraints from strong and weak lensing \citep{Seitz:1997}.
When lensed background sources are available across a wide range of
redshifts (as is the case for the Abell 2744 cluster discussed here),
it should in principle be possible to break this
degeneracy \citep{Seitz:1997,Bradac:2004}.  However, there are more
complex versions of positional constraint
degeneracies \citep{Liesenborgs:2012,Schneider:2014}.  Such
degeneracies do not extend to the {\it absolute} magnification of a
background source's flux and size. In the case of the simple
mass sheet degeneracy described above, the magnification scales as
$\mu\propto(1-\lambda)^{-2}$ \citep[see e.g.][]{Bartelmann:2010}.
Therefore, one can break these fundamental degeneracies with an
absolute measurement of magnification from a standard
candle \citep{Holz:2001} or a standard ruler \citep{Sonnenfeld:2011}.

In Section~\ref{sec:DiscoveryAndFollowup} we present the discovery and
follow-up observations of SN \tomas\ at $z=1.3457$, discovered behind
the galaxy cluster Abell 2744.  Section~\ref{sec:HostGalaxy} examines
the SN host galaxy.  Sections~\ref{sec:Spectroscopy}
and \ref{sec:PhotometricClassification} describe the spectroscopy and
photometry of this SN, leading to a classification of the object as a
normal Type Ia SN.  In Section~\ref{sec:DistanceAndMagnification} we
make a direct measurement of the magnification of this source due to
gravitational lensing.  Section~\ref{sec:Discussion} discusses the
tension between our magnification measurement and the lens models. Our
conclusions are summarized in Section~\ref{sec:SummaryAndConclusions},
along with a discussion of future prospects. 

\section{Discovery, Follow-up, and Data Processing}
\label{sec:DiscoveryAndFollowup}

\begin{figure*}
\begin{center}
\includegraphics[width=\textwidth]{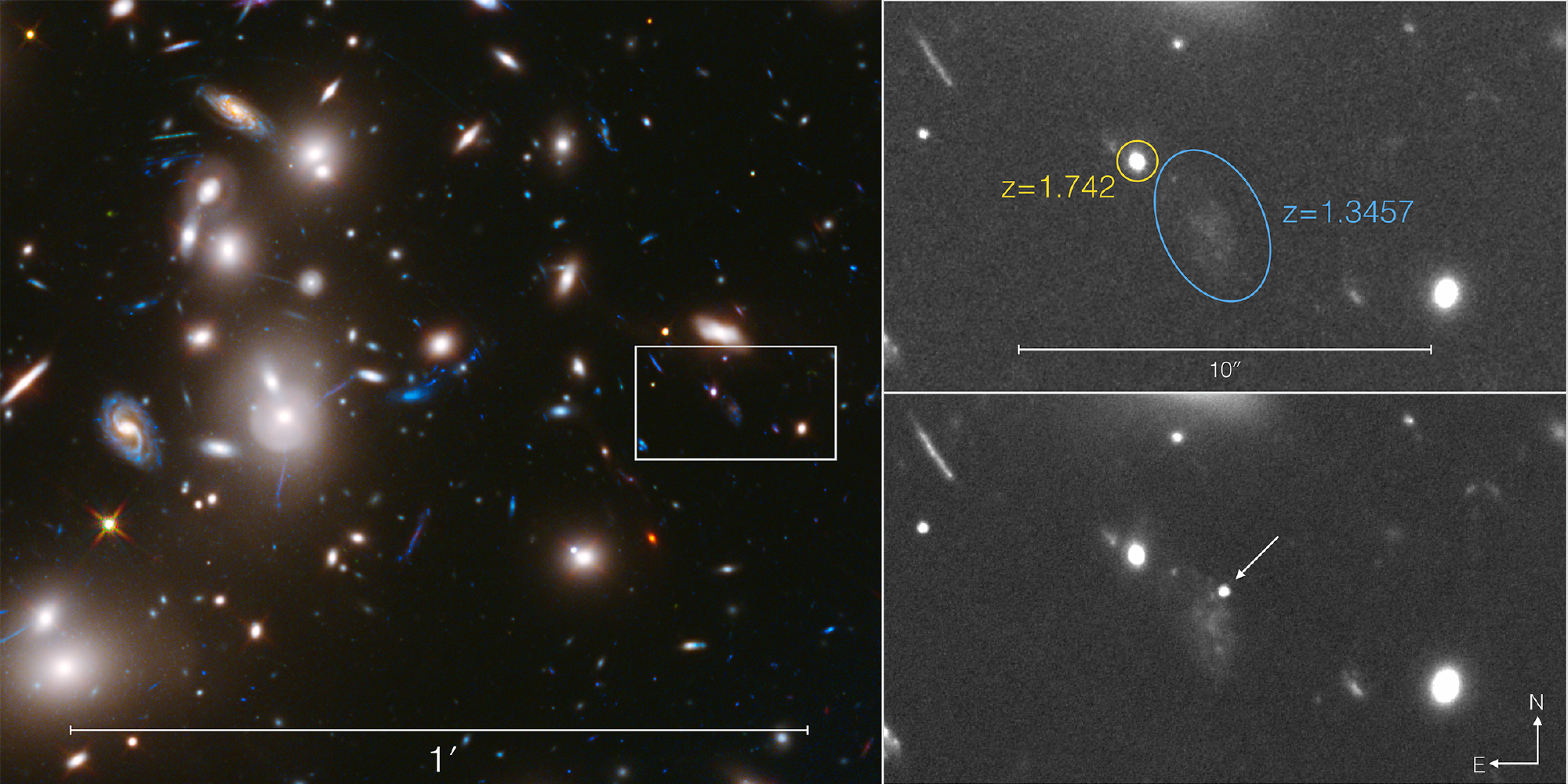}
\caption{  \label{fig:DiscoveryImage} 
SN \tomas\ in the Abell 2744 field.  The left panel shows a
UV/Optical/IR color composite image constructed from all available HST
imaging of the Abell 2744 cluster field.  The inset panels on the right
show F814W imaging of the immediate vicinity of SN \tomas,
approximately 40\arcsec\ from the center of the cluster. The top panel
shows the template image, combining all data prior to the SN
appearance.  Labeled ellipses mark the nearest galaxies and their
spectroscopic redshift constraints, with the most likely host
galaxy marked in blue, and a background galaxy in yellow. The bottom
panel is constructed from all HFF F814W imaging taken while the SN was
detectable, and marks the SN location with an arrow.  (Left panel
image credit: NASA, ESA, and J. Lotz, M. Mountain, A. Koekemoer, and
the HFF Team (STScI)) }
\end{center}
\end{figure*}

SN \tomas\ was discovered in Hubble Space Telescope (\HST) observations
with the Advanced Camera for Surveys (ACS) in the F606W and F814W
bands (V and I), collected on UT 2014 May 15 as part of the Hubble
Frontier Fields (HFF) survey (PI:J.Lotz,
HST-PID:13495).\footnote{\url{http://www.stsci.edu/hst/campaigns/frontier-fields}}
The HFF program is a 3-year Director's discretionary initiative that
is collecting 140 orbits of HST imaging (roughly 340 ksec) on six
massive galaxy clusters, plus 6 accompanying parallel fields.  Each
field is observed in 3 optical bands (ACS F435W, F606W and F814W) and
4 infrared (IR) bands (WFC3-IR F105W, F125W, F140W, and F160W),
although the optical and IR imaging campaigns are separated by $\sim$6
months. Abell 2744 was the first cluster observed, with IR imaging
spanning 2013 October-November, and optical imaging from 2014
May-July.  A composite image of the HFF data showing the SN is
presented in Figure~\ref{fig:DiscoveryImage}, and the
locations of the cluster center, the SN, and the presumed host galaxy
are given in Table~\ref{tab:Positions}.  The SN detection was made in
difference images constructed using template imaging of Abell 2744
from HST+ACS observations taken in 2009 (PI:Dupke, HST-PID:11689).

\begin{deluxetable}{lllll}
\tablecolumns{5}
\tablecaption{J2000 Coordinates of \tomas, host, and cluster. \label{tab:Positions}}
\tablehead{ 
    \colhead{Object}
  & \colhead{R.A.}
  & \colhead{Decl.}
  & \colhead{R.A.}
  & \colhead{Decl.}\\
    \colhead{}
  & \colhead{(h:m:s)}
  & \colhead{(d:m:s)}
  & \colhead{(deg)}
  & \colhead{(deg)}
}
\startdata
\tomas       &  00:14:17.87 &  -30:23:59.7  &  3.574458 &  -30.399917 \\
Host galaxy  &  00:14:17.88 &  -30:24:00.6  &  3.574483 &  -30.400175 \\
Abell 2744\tablenotemark{a}   &  00:14:21.20 &  -30:23:50.1  &  3.588333 &  -30.397250
\enddata
\tablenotetext{a}{Coordinates of the HFF field center, approximately at the center of the cluster.}
\end{deluxetable}

\begin{deluxetable*}{rrlrrlrlcc}
\tablecolumns{10}
\tablecaption{\tomas\ Observations and Photometry\label{tab:Photometry}}
\tablehead{ 
    \colhead{Obs. Date}
  & \colhead{Camera}
  & \colhead{Filter}
  & \colhead{Exp. Time}
  & \colhead{Flux}
  & \colhead{Flux Err}
  & \colhead{AB Mag\tablenotemark{a}}
  & \colhead{Mag Err}
  & \colhead{AB Zero Point} 
  & \colhead{$\Delta$ZP\tablenotemark{b}} \\
    \colhead{(MJD)}
  & \colhead{}
  & \colhead{or grism}
  & \colhead{(sec)}
  & \colhead{(counts/sec)}
  & \colhead{(counts/sec)}
  & \colhead{}
  & \colhead{}
  & \colhead{} 
  & \colhead{(Vega-AB)} 
}
\startdata
56820.06 &  ACS     & F435W &   5083  &    -0.027 &   0.053 &     27.66 & \nodata &   25.665 & -0.102\\        
56821.85 &  ACS     & F435W &   5083  &     0.105 &   0.053 &     28.11 &    0.55 &   25.665 & -0.102\\        
56823.77 &  ACS     & F435W &   5083  &     0.022 &   0.053 &     29.80 &    2.59 &   25.665 & -0.102\\        
56824.97 &  ACS     & F435W &   5083  &     0.021 &   0.053 &     29.85 &    2.72 &   25.665 & -0.102\\        
56828.68 &  ACS     & F435W &   5083  &    -0.148 &   0.053 &     27.65 & \nodata &   25.665 & -0.102\\        
56830.87 &  ACS     & F435W &   5083  &     0.100 &   0.054 &     28.16 &    0.58 &   25.665 & -0.102\\        
56832.86 &  ACS     & F435W &   5083  &    -0.080 &   0.053 &     27.66 & \nodata &   25.665 & -0.102\\        
56833.86 &  ACS     & F435W &   5083  &    -0.002 &   0.053 &     27.67 & \nodata &   25.665 & -0.102\\        
56839.50 &  ACS     & F435W &   5083  &    -0.022 &   0.052 &     27.68 & \nodata &   25.665 & -0.102\\[1mm]   
56792.06 &  ACS     & F606W &   5046  &     0.363 &   0.083 &     27.59 &    0.25 &   26.493 & -0.086\\        
56792.98 &  ACS     & F606W &   3586  &     0.692 &   0.095 &     26.89 &    0.15 &   26.493 & -0.086\\        
56797.10 &  ACS     & F606W &   4977  &     0.968 &   0.087 &     26.53 &    0.10 &   26.493 & -0.086\\        
56800.08 &  ACS     & F606W &   4977  &     0.844 &   0.085 &     26.68 &    0.11 &   26.493 & -0.086\\        
56804.99 &  ACS     & F606W &   5046  &     0.977 &   0.086 &     26.52 &    0.10 &   26.493 & -0.086\\[1mm]   
56792.99 &  ACS     & F814W &   3652  &     1.639 &   0.104 &     25.41 &    0.07 &   25.947 & -0.424\\        
56797.11 &  ACS     & F814W &   4904  &     3.376 &   0.141 &     24.63 &    0.05 &   25.947 & -0.424\\        
56798.95 &  ACS     & F814W &   5046  &     3.951 &   0.156 &     24.46 &    0.04 &   25.947 & -0.424\\        
56800.10 &  ACS     & F814W &   4904  &     3.854 &   0.155 &     24.48 &    0.04 &   25.947 & -0.424\\        
56801.89 &  ACS     & F814W &  10092  &     4.102 &   0.153 &     24.41 &    0.04 &   25.947 & -0.424\\        
56802.95 &  ACS     & F814W &  10092  &     4.325 &   0.160 &     24.36 &    0.04 &   25.947 & -0.424\\        
56803.93 &  ACS     & F814W &  15138  &     4.402 &   0.160 &     24.34 &    0.04 &   25.947 & -0.424\\        
56804.08 &  ACS     & F814W &   5046  &     4.658 &   0.178 &     24.28 &    0.04 &   25.947 & -0.424\\        
56812.08 &  ACS     & F814W &    637  &     4.705 &   0.258 &     24.27 &    0.06 &   25.947 & -0.424\\        
56815.93 &  ACS     & F814W &    446  &     4.026 &   0.285 &     24.43 &    0.08 &   25.947 & -0.424\\        
56820.07 &  ACS     & F814W &   5044  &     3.508 &   0.142 &     24.58 &    0.04 &   25.947 & -0.424\\        
56821.87 &  ACS     & F814W &   5044  &     3.541 &   0.144 &     24.57 &    0.04 &   25.947 & -0.424\\        
56823.79 &  ACS     & F814W &   5044  &     2.876 &   0.124 &     24.80 &    0.05 &   25.947 & -0.424\\        
56824.99 &  ACS     & F814W &   5044  &     3.060 &   0.129 &     24.73 &    0.05 &   25.947 & -0.424\\        
56828.70 &  ACS     & F814W &   5044  &     2.777 &   0.121 &     24.84 &    0.05 &   25.947 & -0.424\\        
56830.89 &  ACS     & F814W &   5044  &     2.395 &   0.111 &     25.00 &    0.05 &   25.947 & -0.424\\        
56832.88 &  ACS     & F814W &   5044  &     2.331 &   0.108 &     25.03 &    0.05 &   25.947 & -0.424\\        
56833.88 &  ACS     & F814W &   5044  &     2.389 &   0.111 &     25.00 &    0.05 &   25.947 & -0.424\\        
56839.52 &  ACS     & F814W &   5044  &     1.673 &   0.093 &     25.39 &    0.06 &   25.947 & -0.424\\[1mm]   
56833.14 &  WFC3-IR & F105W &    756  &     7.504 &   0.239 &     24.08 &    0.03 &   26.269 & -0.645\\        
56841.82 &  WFC3-IR & F105W &    756  &     5.822 &   0.208 &     24.36 &    0.04 &   26.269 & -0.645\\        
56850.06 &  WFC3-IR & F105W &    756  &     3.952 &   0.207 &     24.78 &    0.06 &   26.269 & -0.645\\        
56860.62 &  WFC3-IR & F105W &   1159  &     2.899 &   0.167 &     25.11 &    0.06 &   26.269 & -0.645\\        
56886.63 &  WFC3-IR & F105W &   1159  &     1.216 &   0.147 &     26.06 &    0.13 &   26.269 & -0.645\\        
56891.67 &  WFC3-IR & F105W &    356  &     0.971 &   0.324 &     26.30 &    0.36 &   26.269 & -0.645\\        
56893.20 &  WFC3-IR & F105W &    712  &     0.954 &   0.242 &     26.32 &    0.28 &   26.269 & -0.645\\        
56954.64 &  WFC3-IR & F105W &    356  &     0.521 &   0.388 &     26.98 &    0.81 &   26.269 & -0.645\\[1mm]   
56817.08 &  WFC3-IR & F125W &   1206  &     8.459 &   0.191 &     23.91 &    0.02 &   26.230 & -0.901\\        
56833.15 &  WFC3-IR & F125W &    756  &     7.753 &   0.255 &     24.01 &    0.04 &   26.230 & -0.901\\        
56841.83 &  WFC3-IR & F125W &    806  &     6.015 &   0.227 &     24.28 &    0.04 &   26.230 & -0.901\\        
56850.07 &  WFC3-IR & F125W &    806  &     4.343 &   0.224 &     24.64 &    0.06 &   26.230 & -0.901\\[1mm]   
56891.86 &  WFC3-IR & F140W &    712  &     2.578 &   0.344 &     25.42 &    0.14 &   26.452 & -1.076\\        
56893.06 &  WFC3-IR & F140W &    712  &     3.026 &   0.363 &     25.25 &    0.13 &   26.452 & -1.076\\        
56955.58 &  WFC3-IR & F140W &   1424  &     1.218 &   0.269 &     26.24 &    0.24 &   26.452 & -1.076\\[1mm]   
56817.09 &  WFC3-IR & F160W &   1206  &     4.831 &   0.263 &     24.24 &    0.06 &   25.946 & -1.251\\        
56833.21 &  WFC3-IR & F160W &    756  &     3.965 &   0.241 &     24.45 &    0.07 &   25.946 & -1.251\\        
56841.84 &  WFC3-IR & F160W &    756  &     3.011 &   0.234 &     24.75 &    0.08 &   25.946 & -1.251\\        
56850.08 &  WFC3-IR & F160W &    756  &     2.744 &   0.223 &     24.85 &    0.09 &   25.946 & -1.251\\        
56860.67 &  WFC3-IR & F160W &   1159  &     1.895 &   0.177 &     25.25 &    0.10 &   25.946 & -1.251\\        
56886.64 &  WFC3-IR & F160W &   1159  &     1.677 &   0.191 &     25.38 &    0.12 &   25.946 & -1.251\\[1mm]   
\tableline\\
56812.0 & ACS     & G800L &   3490 &  \nodata & \nodata & \nodata & \nodata & \nodata & \nodata\\
56815.7 & ACS     & G800L &   6086 &  \nodata & \nodata & \nodata & \nodata & \nodata & \nodata
\enddata
\tablenotetext{a}{For non-positive flux values we report the magnitude as a 3-$\sigma$ upper limit}
\tablenotetext{b}{Zero point difference: the magnitude shift for conversion from AB to Vega magnitude units.}
\end{deluxetable*}

Upon discovery, HST target-of-opportunity observations were triggered
from the FrontierSN program (PI:Rodney, HST-PID:13386), which aims to
discover and follow transient sources in the HFF cluster and parallel
fields. The FrontierSN observations provided WFC3-IR imaging as well
as spectroscopy of the SN itself using the ACS G800L grism,
supplementing the rapid-cadence optical imaging from HST+ACS already
being provided by the HFF program. The last detections in the IR F105W
and F140W bands came from the direct-imaging component of the GLASS
program.  Difference images for the IR follow-up data were generated
using templates constructed from the HFF WFC3-IR imaging campaign,
which concluded in November, 2013.

All of the imaging data were processed using the {\tt sndrizpipe}
pipeline,\footnote{\url{https://github.com/srodney/sndrizpipe} v1.2
DOI:10.5281/zenodo.10731} a custom data reduction package in Python
that employs the {\tt DrizzlePac} tools from the Space Telescope
Science Institute (STScI) \citep{Fruchter:2010}.  Photometry was
collected using the {\tt PyPhot} software
package,\footnote{\url{https://github.com/djones1040/PyPhot}} a
pure-Python implementation of the photometry algorithms from the IDL
AstroLib package \citep{Landsman:1993}, which in turn are based on the
DAOPHOT program \citep{Stetson:1987}.  For the IR bands we used point
spread function (PSF) fitting on the difference images, and in the ACS
optical bands we collected photometry with a
0\farcs3 aperture. Table~\ref{tab:Photometry} presents the list of
observations, along with measured photometry from all available
imaging data.

\section{Host Galaxy}
\label{sec:HostGalaxy}

The most probable host galaxy for SN \tomas\ is a faint and diffuse
galaxy immediately to the south-east of the SN location.  With
photometry of the host galaxy collected from the template images, we
fit the spectral energy distribution (SED) using the {\it BPZ} code --
a Bayesian photometric redshift estimator \citep{Benitez:2000}. 
From the {\it BPZ} analysis, we found the host to be most likely an
actively star-forming galaxy at a redshift of
$z=1.5\pm0.2$.  This photo-z was subsequently refined to a
spectroscopic redshift of $z=1.3457\pm0.0001$, based on optical
spectroscopy of the host (Mahler et al. in prep.) that shows two
significant ($>10\sigma$) emission lines at 8740.2 and 8746.7
Angstroms, consistent with the [OII] $\lambda\lambda$ 3726-3729 \AA\
doublet.  

The next nearest galaxy detected in HST imaging is 2.2\arcsec
northeast of the SN position. It has a redshift of $z=1.742$
determined from a spectrum taken with the G141 grism of the HST
WFC3-IR camera, collected as part of the Grism Lens-Amplified Survey
from Space (GLASS, PI:Treu, PID:13459, \citealt{Treu:2015,Schmidt:2014}).  As we
will see in Sections~\ref{sec:Spectroscopy}
and \ref{sec:PhotometricClassification}, both the spectroscopic and
photometric data from the SN itself are consistent with the
redshift of the fainter galaxy at $z=1.3457$, and
incompatible with $z=1.742$ from this brighter galaxy.  This means
that the latter galaxy is a background object and therefore has no
impact on the SN magnification.

The galaxy identified as the host is not close enough to the cluster
core to {\it necessarily} be multiply-imaged, but it is still possible
that the host galaxy is one of the outer images of a multiple image
system.  In such a case, and given the position of the SN host, one
would expect that another image of the galaxy would be present at a
similar brightness, and would therefore be detectable in HST
imaging. To date, no plausible candidate for a counter-image has been
identified.

To measure the stellar mass of the host galaxy
we use Eq. 8 of \citet{Taylor:2011}, which relates the rest-frame
(g-i) color and i-band luminosity to the total stellar mass. To derive
these values, we fixed the redshift at $z=1.3457$ and repeated the SED
fitting using BPZ. From the best-fit SED we extracted rest-frame
optical magnitudes, and corrected them for lensing using a
magnification factor of $\mu=2.0$ -- a value that we will derive from
the SN itself in Section~\ref{sec:DistanceAndMagnification}.  From
this we determine the host galaxy mass to be $10^{9.8}$ \Msun.

\section{Spectroscopy}
\label{sec:Spectroscopy}

\begin{figure}
\begin{center}
\includegraphics[width=\columnwidth]{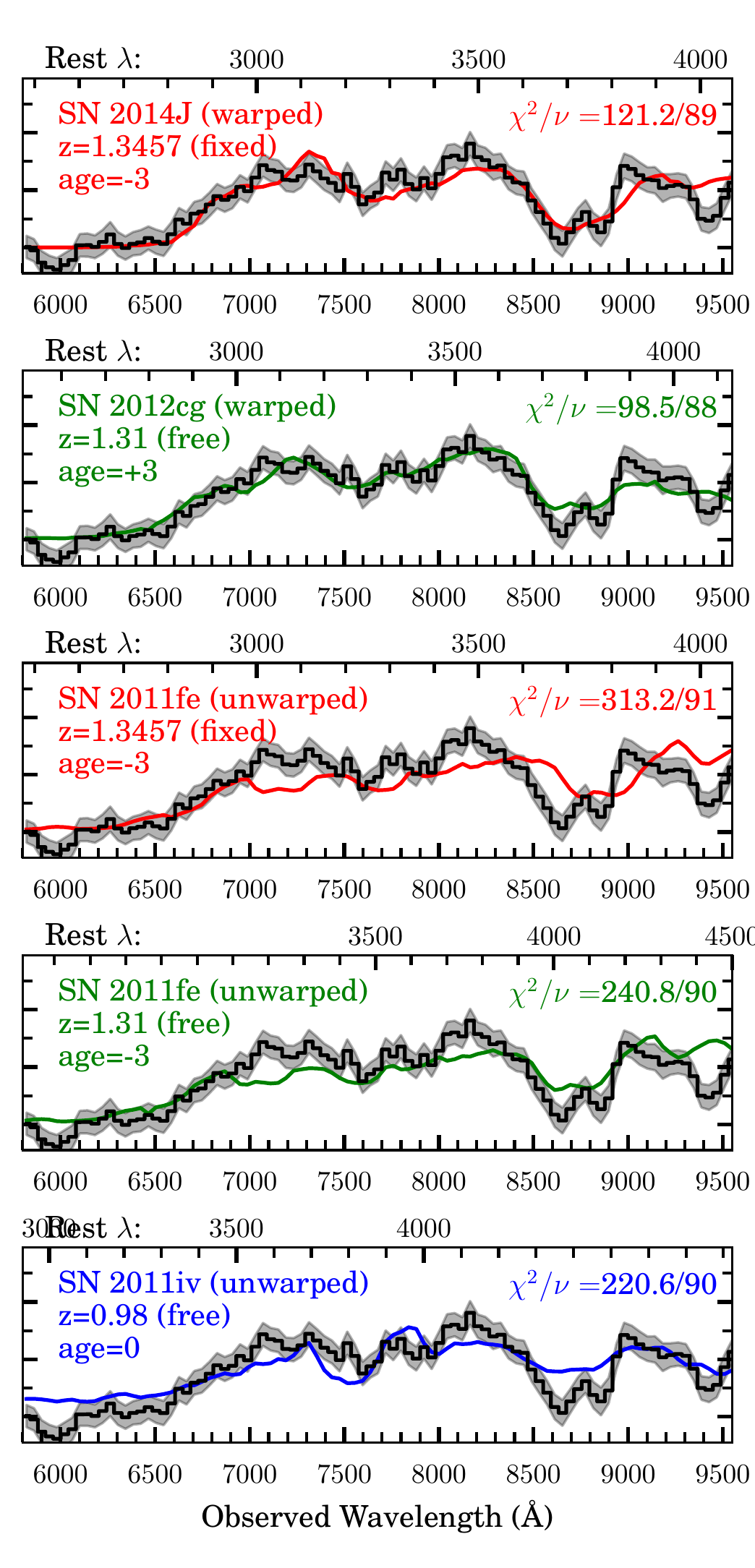}
\caption{  \label{fig:SpecFit}
Redshift and age determination from spectral template matching to the
the SN \tomas\ maximum light spectrum.  The y axis plots flux in
arbitrary units, and the x axis marks wavelength in \AA\ with the
observer-frame on the bottom and rest-frame on the top.  The \tomas\
spectrum observed with the HST ACS G800L grism is shown in black,
overlaid with model fits derived from a library of Type Ia templates
that have extended rest-frame UV coverage.  The top two panels show
the best matching templates when using a smooth 3rd-order polynomial
to warp the shape of the template pseudo-continuum, with the redshift
fixed at z=1.3457 and then allowed to float as a free parameter.  The
lower three panels show matches found when the templates are not
warped, both with and without fixing the redshift.  The bottom panel
shows the best match in this set, although at $z=0.98$ it is
inconsistent with the host galaxy redshift prior and the light curve.
}
\end{center}
\end{figure}

A spectrum of SN \tomas\ was collected with the ACS G800L grism on
2014 June 4 and 7, when the SN was within 3 observer-frame
days of the observed peak brightness in the F814W band.  The
observations -- listed at the bottom of Table~\ref{tab:Photometry} --
used 5 HST orbits from the FrontierSN program for a total
spectroscopic exposure time of $\sim$10 ksec.  The grism data were
processed and the target spectrum was extracted using a custom
pipeline \citep{Brammer:2012}, which was developed for the 3D-HST
program (PI:Van Dokkum; PID:12177, 12328) and also used by the GLASS
team.

Figure~\ref{fig:SpecFit} shows the composite 1-D ACS grism spectrum,
combining all available G800L exposures, overlaid with SN model fits
that will be described below.  The spectrum is largely free of
contamination, because the orientation was chosen to avoid nearby
bright sources and the host galaxy is diffuse and optically faint.
Thus, the SN spectral features can be unambiguously identified, most
notably the red slope of the continuum and a prominent absorption
feature at $\sim$8700\AA.  

Spectra of Type I SNe (including all sub-classes Ia, Ib and Ic) are
dominated by broad absorption features, which cannot in general be
used to directly extract a spectroscopic redshift \citep[see
e.g.][]{Filippenko:1997}.  As described below, we fit template spectra
to the SN \tomas\ data in two steps.  First we determine a spectral
classification -- and get a preliminary estimate of the redshift and
age -- using the SuperNova IDentification (SNID)
software \citep{Blondin:2007}.  Second, we refine the redshift and age
measurement using a custom Type Ia spectral template matching program.

\subsection{Classification with SNID}
\label{sec:SNID}

The SNID program is designed to estimate the type, redshift, and age
of a SN spectrum through cross-correlation matching with a library of
template spectra, using the algorithm of \citet{Tonry:1979}.
To account for possible distortions in the broad shape of the SN
pseudo-continuum due to dust or instrumental calibration effects, SNID
divides each SED by a smooth cubic spline fit. This effectively
removes the shape of the SN pseudo-continuum to leave behind a flat
SED superimposed with spectral absorption and emission features.  It
is these features which drive the cross-correlation fit, so the SNID
approach is insensitive to the overall color of the SED.  We used v2.0
of the SNID template library, which includes template SEDs covering
all Type Ia and core collapse sub-classes, and has recently been
updated with corrections and improvements to the Type Ib/c
templates \citep{Liu:2014}.

In SNID the goodness of fit is evaluated primarily through the {\it
rlap} parameter, which measures the degree of wavelength overlap and
the strength of the cross-correlation peak.  Typically, an {\it rlap}
value $>5$ is required to be considered an acceptable match.  

To match the SN \tomas\ spectrum we use conservative constraints on
age and redshift: limiting the age to $\pm$5 rest-frame days from peak
brightness and $0.8<z<1.8$, consistent with the SN light curve and
the two plausible host galaxies.  With these constraints we find that the only
acceptable match is a normal Type Ia SN near $z=1.3$. The best match
has {\it rlap}$=8.7$, using the normal Type Ia SN 2005cf at $z=1.35$
and age=-2.2 rest-frame days before peak.  In contrast, 
the best non-Ia matches all have {\it rlap}$<2.5$.

Using {\tt SNID} we can find an acceptable \CCSN\ match only when we
remove all age and redshift constraints. In this case the best non-Ia
match is the Type Ic SN 1997ef, which delivers {\it rlap}$=6.8$ at
$z=0.51$ and age=47.3 rest-frame days past peak (71 observer-frame
days).  This is not as good a fit as the best Type Ia models, is at
odds with the host galaxy redshift prior, and is strongly disfavored
by the shape and colors of the SN light curve (see
Section~\ref{sec:PhotometricClassification}).

From the preceding analysis, we conclude that \tomas\ is a Type Ia SN
at $z\approx1.35$.  At this redshift, the absorption at $\sim$8700\AA\
corresponds to the blended \ion{Ca}{2} H\&K and \ion{Si}{2} $\lambda$3858 features.  Generally referred to as the Ca H\&K feature, this absorption is commonly seen in Type Ia SN spectra
near maximum light, although it is also prominent in the spectra of
Type Ib and Ic \CCSNe.  The red color of
the \tomas\ SED is qualitatively consistent with a redshift of $z>1$
-- although this information was not used by {\tt SNID} for the
template matching.  As we will see in
Section~\ref{sec:PhotometricClassification}, this spectral
classification of SN \tomas\ is reinforced by the photometric
information, which also supports classification as a Type Ia SN at
$z\approx1.35$.

\subsection{Spectral Fitting with UV Type Ia Templates}
\label{sec:SpecRedshift}

To refine the redshift and phase constraints on \tomas, we next fit
the spectrum with a custom spectral matching program that employs a
library of Type Ia SN SEDs.  This library is similar to the Type Ia
spectral set used by SNID, but also includes more recent SNe with
well-observed spectral time series that extend to rest-frame UV
wavelengths (e.g. SN 2011fe and 2014J).  We first use an approach
similar to the SNID algorithm: warping the pseudo-continuum of each
template spectrum by dividing out a 3$^{\rm rd}$-order polynomial to
match the observed SED of SN \tomas.  This approach will find
templates that have similar abundances and photospheric
velocities. With the redshift fixed at $z=1.3457$ we find the best fit
is a spectrum from the normal Type Ia SN 2014J \citep{Foley:2014},
with a $\chi^2$ per degree of freedom $\nu$ of $\chi^2/\nu=121.8/89$
shown in the top panel of Figure~\ref{fig:SpecFit}. The excess
variance in this fit may be attributed to the intrinsic variation of
Type Ia SN spectra, which is more prominent at UV
wavelengths \citep[e.g.][]{Foley:2008b,Wang:2012} and is not fully
represented in the available template library.  Allowing the redshift
as an additional free parameter, we still find results that are
consistent with the SNID fits and $<2\sigma$ from the host galaxy
redshift: $z=1.31\pm0.02$ and a phase of $0\pm3$ rest-frame days. The
best-fitting spectral template in this case is the normal Type Ia SN
2012cg \citep{Amanullah:2015} at $z=1.31$ with $\chi^2/\nu=98.5/88$
(second panel of Figure~\ref{fig:SpecFit}).

Next, we repeat the fitting, but without any warping of the templates
to account for differences in the continuum shape.  In this iteration
we only allow each template SED to be scaled in flux coherently at all
wavelengths, so the fits are more sensitive to the overall color of
the SED.\footnote{Note that gravitational lensing does not affect the color of background sources at all, so the expected lensing magnification of \tomas\ does not affect this analysis.}  Fixing the redshift to $z=1.3457$, we find the best match is
from the normal Type Ia SN 2011fe \citep{Mazzali:2014}, though the fit is quite poor, with
$\chi^2/\nu=313.2/91$ (third panel of Figure~\ref{fig:SpecFit}).  When
the redshift is allowed as a free parameter the \tomas\ SED is still
matched best by a SN 2011fe template, now at redshift $z=1.31$ with
$\chi^2/\nu=240.8/90$ (fourth panel of Figure~\ref{fig:SpecFit}).  
SN 2011fe had effectively no dust reddening \citep[e.g.][]{Nugent:2011,Li:2011c}, so the fact that SN 2011fe provides the best un-warped template match is further evidence that SN \tomas\ suffers from very little dust extinction.

Without the continuum warping, an alternative fit also arises: the
fast-declining (91bg-like) Type Ia SN 2011iv \citep{Foley:2012b} at
$z=0.98\pm0.01$. Formally, this match provides a slightly better fit
to the unwarped \tomas\ spectrum ($\chi^2/\nu=220.6/90$,
bottom panel of Figure~\ref{fig:SpecFit}), although the fit is
notably poorer at $\sim8700$\AA\ where the most significant absorption
feature is found. Furthermore, a redshift $z\sim1$ is at odds with the
spectroscopic redshift of the nearest galaxy ($z=1.3457$), and we will
see in the following section that the photometric data is also
incompatible with a Type Ia SN at $z\sim1.0$.

Setting aside the $z\sim1$ solution, all other template matches
provide a consistent redshift constraint of $z=1.31\pm0.02$,
regardless of whether the templates are warped to match the SN \tomas\
continuum shape.  The inferred age from these fits is $0\pm3$
rest-frame days from peak brightness, which is also consistent with
the observed light curve.  Taken together with the host galaxy
spectroscopic redshift, these fits suggest that SN \tomas\ is a normal
Type Ia SN at $z=1.3457$ with an SED color close to SN 2011fe, but
spectral absorption features similar to SN 2014J.

\section{Photometric Classification}
\label{sec:PhotometricClassification}

Relative to other SNe at $z>1$, the SN \tomas\ light curve was
unusually well sampled at rest-frame ultraviolet wavelengths, due to
the rapid cadence of the HFF imaging campaign. These ACS observations
therefore provide a tight constraint on the time of peak brightness
and the evolution of the SN color.  Supplemental observations with the
WFC3-IR camera provided critical rest-frame optical photometry,
enabling a measurement of the apparent luminosity distance through
light curve fitting.

\begin{figure}
\begin{center}
\includegraphics[width=\columnwidth]{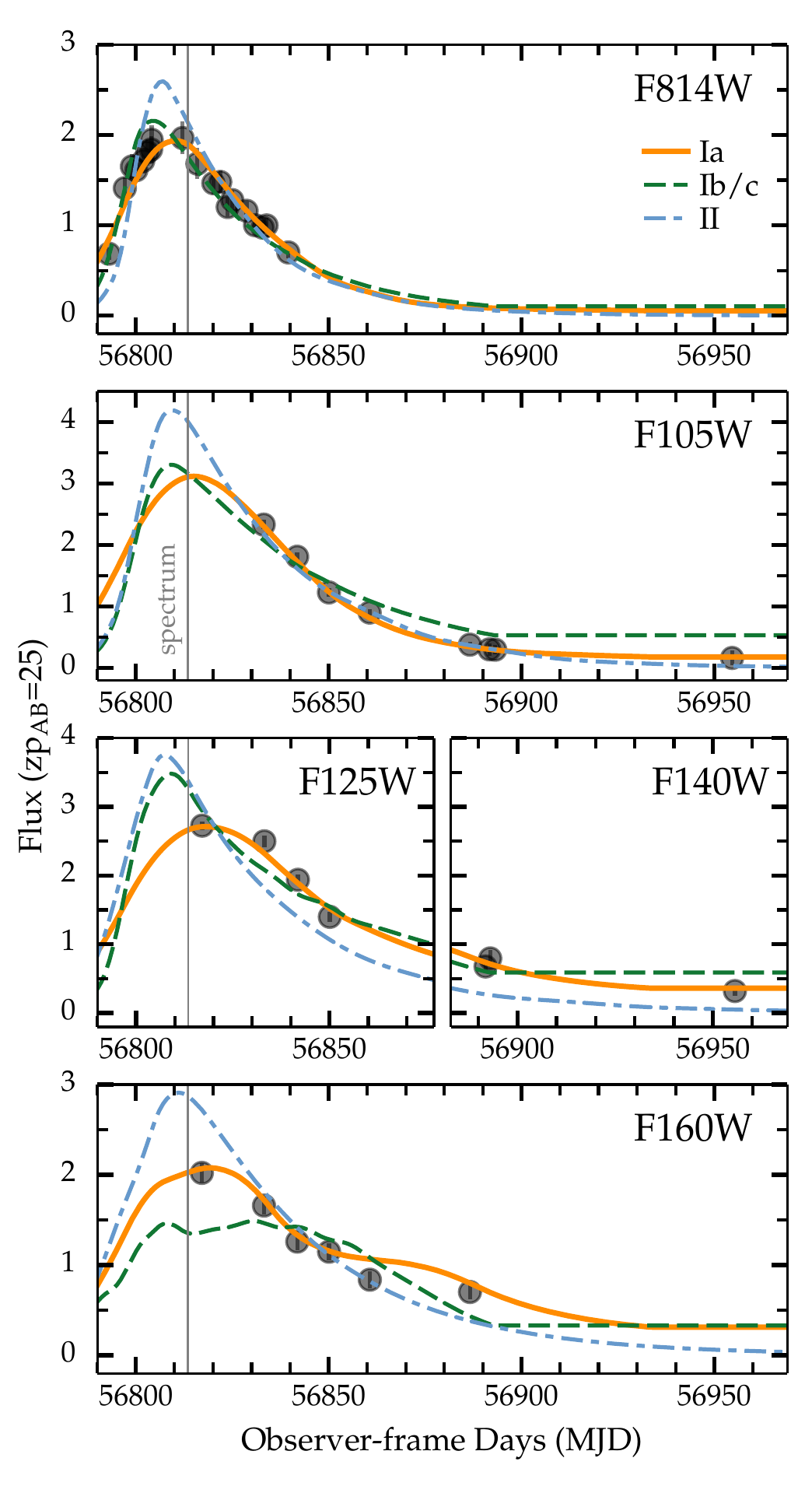}
\caption{ 
Maximum likelihood model for each SN sub-class, derived from Bayesian
model selection using the photometric data alone. Grey points show the
observed SN \tomas\ photometry with error bars, though these are
typically smaller than the size of the marker.  The Type Ia model
(orange solid line) is drawn from the SALT2 template at $z=1.35$. The
best match from all Type Ib/Ic models is based on the Type Ic SN
SDSS-14475 at $z=0.695$ (green dashed line). For the Type II class,
the best match is from the Type II-L SN 2007pg at $z=1.8$ (blue
dash-dot line). The Type Ia model is by far the best match, and the
only one that is consistent with both the \change{spectroscopic
redshift} of the probable host galaxy and the spectroscopic redshift
from the SN spectrum.  The date of the HST spectral observations is
marked with a thin grey vertical line.
\label{fig:photoclass} }
\end{center}
\end{figure}

As a check on the spectral classification of SN \tomas\
(Section~\ref{sec:SNID}), we independently classified the SN using a
Bayesian photometric classifier.  We use the {\tt sncosmo} software
package\footnote{\url{http://sncosmo.github.io/}} to simulate SN light
curves from $z=0.3$ to 2.3 and evaluate the classification probability
using traditional Bayesian model selection \citep[as
in][]{Jones:2013,Rodney:2014,Graur:2014,Rodney:2015b}.  In this
analysis we represent normal Type Ia SNe with the SALT2
model \citep{Guy:2010}, and \CCSNe\ with 42 discrete templates (26
Type II and 16 Type Ib/c) drawn from the template library of the
SuperNova Analysis software
package \citep[SNANA,][]{Kessler:2009a}.\footnote{Throughout this work
we use SNANA v10\_35g.}  Likelihoods are defined by comparing the
observed fluxes to model predictions in all passbands where the model
is defined.  In practice, this means we exclude the SN detections in
the F435W and F606W bands, which are too blue for our models at
$z>0.85$.

The \CCSN\ models have free parameters for date of peak brightness
($t_{\rm pk}$), amplitude, and redshift ($z$). Due to the expected
impact of gravitational lensing magnification, we \change{use a flat
prior for the intrinsic luminosity of all SN sub-classes.  We also
apply a flat prior for the SN redshift, which} allows our photometric
analysis to provide an independent check on the host galaxy photo-$z$
and spectroscopic redshift (Sections~\change{\ref{sec:HostGalaxy}}
and \ref{sec:SpecRedshift}).

For Type Ia SNe, the SALT2 model has two additional parameters that
control the shape ($x_1$) and color ($c$) of the light curve.  We use
conservative priors here, defined to encompass a range of Type Ia SN
shapes and colors that is broader than typically allowed in
cosmological analyses \citep[see
e.g.,][]{Kessler:2009b,Sullivan:2011,Rest:2014}. For $x_1$ the prior
is a bifurcated Gaussian distribution with mean $\bar{x}_1=0$,
dispersion $\sigma_{x_1}^+={0.9}$ and $\sigma_{x_1}^-={-1.5}$.  The
bifurcated Gaussian prior for the color parameter $c$ has
$\bar{c}=0.0$, $\sigma_c^-=0.08$, and $\sigma_c^+=0.54$.  The $c$
parameter in SALT2 combines intrinsic SN color and extinction due to
dust, so the large red tail of this distribution allows for the
possibility of several magnitudes of dust extinction along the \tomas\
line of sight.

We also assign a class prior for each of the three primary SN
sub-classes (Type Ia, Ib/c, and II), using a fixed relative fraction for
each sub-class as determined at $z=0$ by \citet{Smartt:2009}
and \citet{Li:2011a}.  A more rigorous classification would extrapolate
these local SN class fractions to higher redshift using models or
measurements of the volumetric SN rate.  For simplicity, we do not
vary the class priors with redshift, and in practice these priors do
not have any significant impact on the resulting classification.

The final photometric classification probability for SN \tomas\ is
$p({\rm Ia}|{\rm {\bf D}})=1.0$, with the classification probability
from all \CCSN\ sub-classes totaling less than
$10^{-32}$.  Although this Bayesian classification utilizes
the full posterior probability distribution, for illustration we
highlight in Figure~\ref{fig:photoclass} a single best-fit model for
each sub-class. This demonstrates how the \CCSN\ models fail to
adequately match the observed photometry. In particular, only the Type
Ia model can simultaneously provide an acceptable fit to the
well-sampled rising light curve in F814W and the F814W-F160W color
near peak.

The marginal posterior distribution in redshift for the Type Ia model
is sharply peaked at $z=1.35\pm0.02$, which is fully
consistent with the redshift of the presumed host galaxy ($z=1.3457$)
as well as the redshift of $z=1.31\pm0.02$ derived from the SN
spectrum in Section~\ref{sec:SpecRedshift}.  The time of peak
brightness is also tightly constrained at $t_{\rm pk}=56816.3\pm0.3$,
which means the HST grism observations were collected within 2
rest-frame days of the epoch of peak brightness -- also consistent
with our spectroscopic analysis.

\section{Distance Modulus and Magnification}
\label{sec:DistanceAndMagnification}

With the type and redshift securely defined as a normal Type Ia SN at
$z=1.3457$, we now turn to fitting the light curve with Type Ia
templates to measure the distance
modulus (\S\ref{sec:LightCurveFitting}) and then derive the
gravitational lensing
magnification (\S\ref{sec:ControlSampleComparison}).  In this
process, we would like to avoid introducing systematic uncertainties
inherent to any assumed cosmological model.  To that end, in
section \ref{sec:ControlSampleComparison} we
follow \P14\ and define the magnification by comparing
the measured distance modulus of SN \tomas\ against an average
distance modulus derived from a ``control sample'' of unlensed Type Ia
SNe at similar redshift. This allows us to make only the minimal
assumption that the redshift-distance relationship for Type Ia SN is
smooth and approximately linear over a small redshift span, which
should be true for any plausible cosmological model.

\begin{figure*}
\begin{center}
\includegraphics[width=\textwidth]{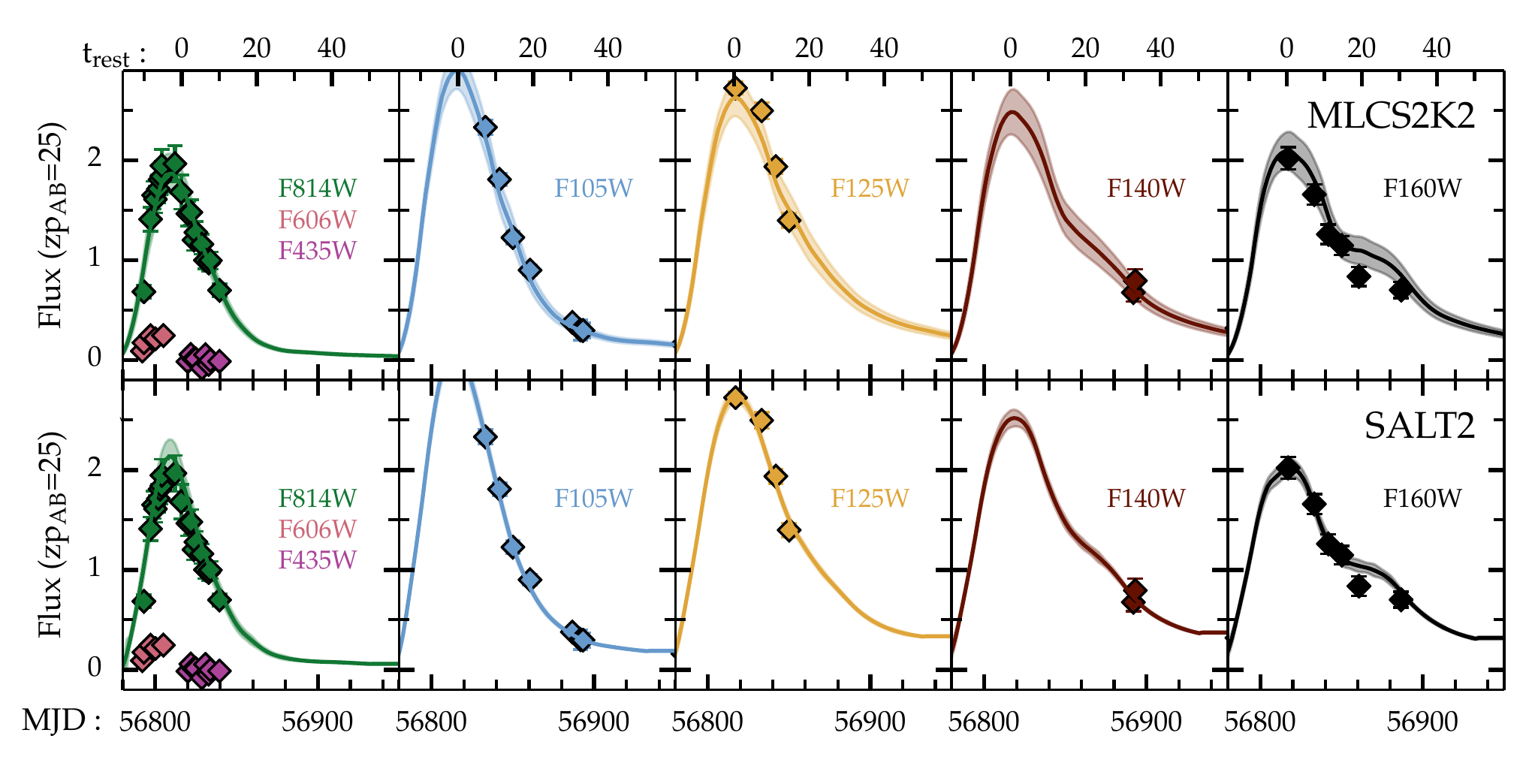}
\caption{ Type Ia light curve fits to SN \tomas\ using the MLCS2k2
(top row) and SALT2 (bottom row) fitters. The model redshifts are set to
$z=1.3457$ as determined from combined
spectroscopic and photometric constraints.  Solid lines denote the
best-fit model and shaded lines show the range allowed by 1-$\sigma$
uncertainties on the model parameters. 
Observed fluxes are shown as diamonds, scaled to an AB magnitude
zero point of 25. Error bars are plotted, but most are commensurate
with the size of the points. The left-most panel includes observations
in the F435W and F606W filters, although these were not used for the
fit, as they are bluer than the minimum wavelength for the model. The
lower axis marks time in observer frame days, while the top axis shows
the time in the rest frame relative to the epoch of peak brightness.
\label{fig:LightCurveFits} }
\end{center}
\end{figure*}

\subsection{Light Curve Fitting}
\label{sec:LightCurveFitting}

As in \P14, we derive a distance modulus for
SN \tomas\ and all SNe in our control sample using two independent
light curve fitters: the SALT2 model described above and the MLCS2k2
model \citep{Jha:2007}.    With both fitters we find light curve
shape and color parameters for \tomas\ that are fully consistent with
a normal Type Ia SN.  For SALT2, with the
redshift fixed at $z=1.3457$, we find a light curve shape
parameter of $x_1=0.135\pm0.199$ and a color parameter of
$c=-0.127\pm0.025$, yielding a $\chi^2$ value of 45.0 for 36 degrees
of freedom, $\nu$.  With the MLCS2k2 fitter the best-fit shape
parameter is $\Delta=-0.082\pm0.070$ and the color term is
$A_{\rm V}=0.011\pm0.025$, giving $\chi^2/\nu=22.8/36$.

 The MLCS2k2 fitter returns a distance
modulus\footnote{We use 'dm' to indicate the distance modulus to avoid
confusion, reserving the symbol $\mu$ to refer to the lensing
magnification. This 'dm' is a standard distance modulus, defined as
${\rm dm}=5\log_{10}d_L+25$, where $d_L$ is the luminosity distance in
Mpc.}  $\rm{dm}_{\rm MLCS2k2}$ directly, as it is defined to be one of
the free parameters in the model. To derive dm from the SALT2 fit, we
use

\begin{equation} \label{eqn:dmSALT2}
 {\rm dm}_{\rm SALT2} = m_B^* - M + \alpha(s-1) - \beta C.
\end{equation}

\noindent  Here the parameters for light curve shape $s$ and color $C$ 
correspond to the SiFTO light curve fitter \citep{Conley:2008}, so we
first use the formulae from \citet{Guy:2010} to convert from SALT2
($x_1$ and $c$) into the equivalent SiFTO parameters.  We also add an
offset of 0.27 mag to the value of $m_B^*$ returned by SNANA, in order
to match the arbitrary normalization of the SALT2 fitter used
by \citet{Guy:2010} and \citet{Sullivan:2011}. This conversion from SALT2 to SiFTO is necessary, as it allows
us to adopt values for the constants $M$, $\alpha$, and $\beta$
from \citet{Sullivan:2011}, which have been calibrated using 472 SNe
from the SNLS3 sample \citep{Conley:2011}: $M=-19.12\pm0.03$,
$\alpha=1.367\pm0.086$, and $\beta=3.179\pm0.101$.

\begin{deluxetable}{lccc}
\tablecolumns{4}
\tablecaption{\tomas\ Measured Distance Modulus and Magnification at $z=1.3457$\label{tab:MeasuredMagnification}}
\tablehead{ 
    \colhead{}
  & \multicolumn{2}{c}{Distance Modulus}
  & \colhead{Measured}\\
    \colhead{Fitter}
  & \colhead{\tomas}
  & \colhead{Control}
  & \colhead{Magnification}
}
\startdata
MLCS2k2  & $44.205\pm0.12$ & $44.97\pm0.06$ & $2.03\pm0.29$\\
SALT2    & $44.177\pm0.18$ & $44.92\pm0.06$ & $1.99\pm0.38$
\enddata
\end{deluxetable}

The SNANA version of the MLCS2k2 fitter returns a value for the
distance modulus (dm$_{\rm MLCS2k2}$) that has an arbitrary zero point
offset relative to the SALT2 distances (dm$_{\rm SALT2}$).  To put the
two distances onto the same reference frame we add a zeropoint
correction of 0.20 mag to the MLCS2k2 distances as
in \P14.  This correction was derived by applying both
fitters to a sample of Type Ia SNe from the SDSS
survey \citep{Holtzman:2008,Kessler:2009b}, with the extinction law
$R_V$ fixed at 1.9.

The total uncertainty in the distance modulus is  

\begin{equation} \label{eqn:sigmatot}
 \sigma_{\rm tot} = \sqrt{ \sigma_{\rm stat}^2 + \sigma_{int}^2}.
\end{equation}

\noindent
The $\sigma_{\rm stat}$ term is the statistical uncertainty, which
encapsulates uncertainties from the data and the model, and
$\sigma_{\rm int}$ accounts for the remaining {\it unmodeled} scatter.
This latter term is derived by finding the amount of additional
distance modulus scatter that needs to be added to a Type Ia SN
population to get $\chi^2$ per degree of freedom equal to 1 for a
fiducial cosmological model fit to the SN Ia Hubble diagram (distance
modulus vs redshift).  Thus, $\sigma_{\rm int}$ is designed to account
for any unknown sources of scatter in the Type Ia SN population,
including unidentified errors in the data analysis as well as the
natural scatter in intrinsic Type Ia SN luminosities.  The value of
$\sigma_{\rm int}$ may be expected to vary as a function of redshift
and also from survey to survey.

Recent work has highlighted the inadequacy of this simplistic approach
for handling intrinsic scatter in the Type Ia SN
population \citep{Marriner:2011,Kessler:2013,Mosher:2014,Scolnic:2014a,Betoule:2014},
but a full consideration of those alternative approaches is beyond the
scope of this work.  We adopt the simple approach of using a single
empirically defined value for $\sigma_{\rm int}$, reflecting
principally an intrinsic scatter in Type Ia SN intrinsic luminosity
that is fixed across time and phase.  Measurements of $\sigma_{\rm
int}$ range from 0.08 mag \citep{Jha:2007,Conley:2011} to 0.15
mag \citep{Kessler:2009b,Suzuki:2012}.  We adopt a value of
$\sigma_{\rm int}=0.08$ mag, as derived by \citet{Jha:2007}
and \citet{Conley:2011}.  Although this is on the low end of the range
reported in the literature, this value is the most appropriate to
apply to our analysis for two reasons.  First and foremost, for the
SALT2 fitter we are using light curve fit parameters
($\alpha$,$\beta$) with associated uncertainties that have been
derived from the joint analysis of \citet{Conley:2011}
and \citet{Sullivan:2011}. Similarly, the implementation of MLCS2k2
that we have used is based on the uncertainty model derived from model
training in \citet{Jha:2007}.  Inflating $\sigma_{int}$ beyond 0.08
mag would therefore be equivalent to driving the reduced $\chi^2$ of
the Type Ia SN Hubble diagram to $<1$.  Second, the value of 0.08 mag
determined in \citet{Conley:2011} is specific to the HST SN sample,
from \citet{Riess:2007} and \citet{Suzuki:2012}, which is the SN
subset that is most similar to \tomas\ in terms of redshift and data
analysis.  Larger values for $\sigma_{\rm int}$ are typically
associated with SN samples at significantly lower redshifts that have
less homogeneous data collection and analysis. 

Final values for the SN distance modulus are shown in
Table~\ref{tab:MeasuredMagnification}, adopting the spectroscopic
redshift ($z=1.3457$).  In addition to modifying the distance modulus
uncertainty for SN \tomas, we also add $\sigma_{\rm int}=0.08$ mag in
quadrature to the uncertainty for every SN in the control sample. Note
however that the uncertainty on the control sample value at $z=1.3457$
is only 0.06 mag, smaller than the intrinsic dispersion of
any single SN because it reflects our measurement error on the mean
distance modulus of the population.  The distance moduli derived from
the SALT2 and MLCS2k2 light curve fitters are fully consistent within
the uncertainties. 

\subsubsection{Host Galaxy Mass Correction}
\label{sec:HostGalaxyMassCorrection}

It is now an accepted practice in cosmological analyses using Type Ia
SN to apply a correction to the luminosity of each SN based on the
stellar mass of its host galaxy. Typically this is described as a
simple bifurcation of the SN population: SNe that appear in more
massive hosts are observed to be $\sim$0.08 mag brighter (after
corrections for light curve shape and color) than SNe in low-mass
hosts \citep{Kelly:2010,Sullivan:2010}.  The dividing line for this
purely empirical ``mass step'' correction is generally set around
$10^{10}$\Msun.  Although this threshold value is somewhat arbitrary \citep[see
e.g.][]{Betoule:2014}, it happens to be very close to the
-SN \tomas\ host galaxy mass of $10^{9.8}$\Msun (Section~\ref{sec:HostGalaxy}).

The physical mechanism that drives the mass step is not yet
understood, but may be related to the metallicity or age of the SN
progenitor systems.  In either case, the significance of this effect
should decrease with redshift, as metal-rich passive galaxies become
much less common at $z>1$ \citep[see
e.g.][]{Rigault:2013,Childress:2014b}.  Indeed, when the size of the
mass step correction is allowed to vary with redshift, there is no
significant evidence that a non-zero mass step is required at
$z>1$ \citep{Suzuki:2012,Shafer:2014,Betoule:2014}.  Given the
absence of a clear physical model and the lack of empirical support
for a high-$z$ mass step, we do not apply any correction to SN \tomas\
or the control sample. 

If we were to apply the correction to this sample using the standard
approach, then the SN \tomas\ distance modulus would not be adjusted,
because its host galaxy mass (Section~\ref{sec:HostGalaxy}) is just
below 10$^{10}$ \Msun. Note that there is some circularity here, as we
used the SN magnification in deriving the host galaxy mass, and now
use the host mass to inform the magnification.  However, regardless of
whether the \tomas\ host mass falls above the mass threshold or below,
this correction is not substantial enough to account for the observed
magnification tension.  The appropriate correction to be applied to
a SN above the mass threshold in this redshift range would be $<$0.04
mag \citep{Rigault:2013,Childress:2014b}, which results in a change to
the inferred magnification of \tomas\ that is much less than the
observed discrepancy of $\sim$0.24 mag.

\subsection{Control Sample Comparison}
\label{sec:ControlSampleComparison}

\begin{figure}
\begin{center}
\includegraphics[width=\columnwidth]{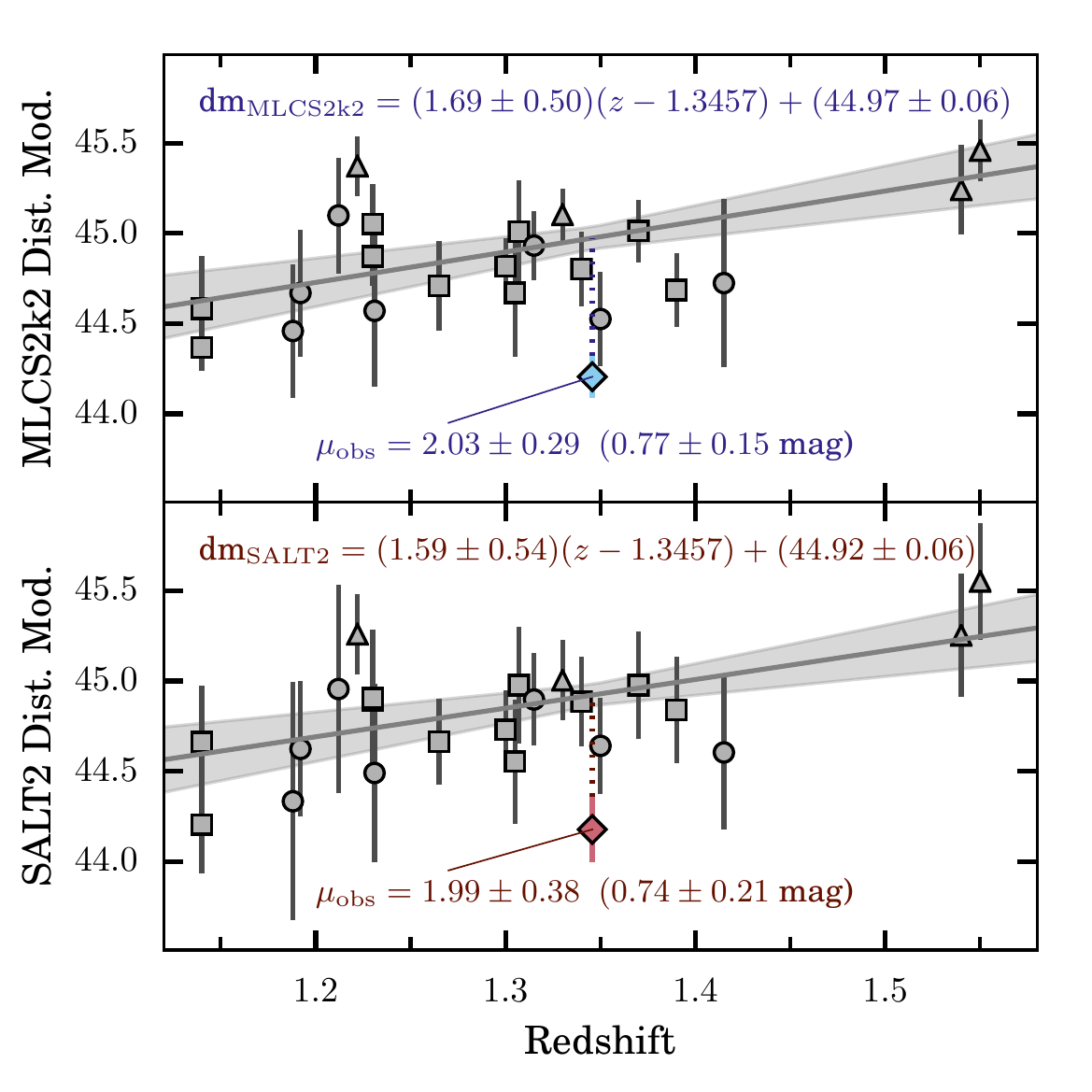}
\caption{ Measurement of the lensing magnification from comparison of
the \tomas\ distance modulus to a sample of unlensed field SN.  SNe
from the GOODS program are plotted as squares, those from the SCP
survey are shown as circles, and the CANDELS objects are triangles.
The distance modulus for each SN is derived from light curve fits
using the MLCS2k2 fitter (top panel) and the SALT2 fitter (bottom
panel).  Grey lines with shading show a linear fit to the unlensed
control sample is shown, anchored at the redshift of SN \tomas, with
fit parameters given at the top.  The derived magnification is
reported at the bottom of each panel.
\label{fig:MagnificationMeasurement} }
\end{center}
\end{figure}

The unlensed sample comprises 22 spectroscopically confirmed Type Ia
SNe in the range $1.1<z<1.6$ from three HST surveys: 11 from the GOODS Higher-$z$ SN search\footnote{GOODS: the Great Observatories Origins Deep Survey, PI:Giavalisco, HST-PID:9425,9583} \citep{Strolger:2004,Riess:2007}, 7 from SCP\footnote{SCP: the Supernova Cosmology Project, PI:Perlmutter} \citep{Suzuki:2012}, and 4 from CANDELS\footnote{CANDELS: the Cosmic Assembly Near-infrared Deep Extragalactic Legacy Survey, PI:Faber \&\ Ferguson} \citep{Rodney:2012,Rodney:2014}.  Using the SALT2 and MLCS2k2
fitters as described above, we get distance modulus measures for every
object in this control sample.  We then fit a linear relationship for
distance modulus vs. redshift, and derive a prediction for the
distance modulus of a normal Type Ia SN at the redshift of SN \tomas\
(Figure~\ref{fig:MagnificationMeasurement}).
This predicted value is given in
Table~\ref{tab:MeasuredMagnification} under the ``Control'' column.
The difference between the observed distance of SN \tomas\ and this
control sample value is attributed to the magnification from
gravitational lensing:

\begin{equation} \label{eqn:mu}
{\rm dm}_{\rm control} - {\rm dm}_{\rm \tomas} = 2.5 \log_{10}\mu.
\end{equation}

\noindent The inferred magnifications for the two different fitters are
reported in the final column of Table~\ref{tab:MeasuredMagnification}.
Although we have chosen to use a cosmology-independent approach to
determine the magnifications, we note that these results are fully
consistent with the values that would be determined by comparing the
observed SN \tomas\ distance modulus against the predicted value from
a flat \LCDM\ cosmology. Adopting the cosmological parameters used
in \P14\ and derived in \citet{Sullivan:2011}
(\Ho=71.6, \Om=0.27, \OL=0.73) would give a predicted distance modulus
${\rm dm}_{\Lambda CDM}=44.90$ and a measured magnification $\mu=1.9$.
Substituting alternative cosmological
parameters \citep[e.g.,][]{Betoule:2014,Planck:2015} would not
significantly change the inferred magnification or affect our
conclusions.

\section{Comparison to Model Predictions}
\label{sec:Discussion}
\label{sec:ComparisonToModelPredictions}

Before the Frontier Fields observations began, the Space Telescope
Science Institute (STScI) issued a call for lens modeling teams to
generate mass models of all 6 Frontier Field clusters, using a shared
collection of all imaging and spectroscopic data available at the
time.  In response to this opportunity, five teams generated seven
models for Abell 2744.  These models necessarily relied on
pre-HFF data, and were required to be complete before the HFF program
began, in order to enable the estimation of magnifications for any new
lensed background sources revealed by the HFF imaging. An interactive
web tool was created by D. Coe and hosted at STScI, to extract
magnification estimates and uncertainties from each model for any
given redshift and position.  In this work we also consider 8
additional models that were created later.  Some of these are
updates of the original models produced in response to the lens
modeling call,  and some of them include new
multiply-imaged galaxies discovered in the HFF imaging as well as new
redshifts for lensed background galaxies.  Magnifications derived from
several of these later models are also available on the interactive
web tool hosted by STScI.  The full list of models and details on
their construction are given in Table~\ref{tab:LensModels}.

\begin{deluxetable*}{p{0.62in}rrrrp{1.4in}p{2.5in}}
\tablecolumns{7}
\tablecaption{Tested lens models for Abell 2744.\label{tab:LensModels}}
\tablehead{ \colhead{Model} & \colhead{N$_{\rm sys}$\tablenotemark{a}} & \colhead{N$_{\rm im}$\tablenotemark{b}} & 
            \colhead{N$_{\rm spec}$\tablenotemark{c}} & \colhead{N$_{\rm phot}$\tablenotemark{d}} & 
            \colhead{References} & \colhead{Description}}
\startdata
Bradac(v1)\tablenotemark{e}     & 16 &  56 & 2 & 14 &  HFF\tablenotemark{f}; \citealt{Bradac:2009} & {\tt SWUnited}\tablenotemark{g} : Free-form, strong+weak-lensing based model.Errors from bootstrap resampling only weak-lensing constraints.\\
CATS(v1)       & 17 &  60 & 2 & 14 &  HFF; \citealt{Richard:2014} &  CATS\tablenotemark{h} team implementation of LENSTOOL\tablenotemark{i} parametric strong-lensing based model.\\
CATS(v1.1)\tablenotemark{j}     & 17 &  60 & 2 & 14 &  \citealt{Richard:2014} &  CATS team LENSTOOL parametric model with both strong and weak lensing constraints.\\
Merten\tablenotemark{e}         & 16 &  56 & 2 & 14 &  HFF; \citealt{Merten:2011} &  {\tt SaWLENS},\tablenotemark{k} Grid-based free-form strong+weak lensing based model using adaptive mesh refinement.\\
Sharon(v1)     & 17 &  60 & 2 & 14 &  HFF;  & LENSTOOL parametric strong-lensing based model\\
Sharon(v2)     & 15 &  47 & 3 & 11 &  \citealt{Johnson:2014} & LENSTOOL parametric strong-lensing based model. Includes cosmological parameter variations in uncertainty estimates.\\
Zitrin-LTM     & 10 &  44 & 2 & 0  &  HFF; \citealt{Zitrin:2009a} & Parametric strong-lensing model, adopts the Light-Traces-Mass assumption for both the luminous and dark matter.\\
Zitrin-NFW     & 10 &  44 & 2 & 0  &  HFF; \citealt{Zitrin:2013a} &  Parametric strong-lensing model using PIEMD profiles for galaxies and NFW profiles for dark matter halos.\\
Williams       & 10 &  40 & 2 & 8  &  HFF & {\tt GRALE}\tablenotemark{l} : Free-form strong-lensing model using a genetic algorithm.  \\
\cutinhead{Post-HFF models : include data from the HFF program}\\
CATS(v2)       & 50 & 151 & 4 & 1  &  \citealt{Jauzac:2015b} & Updated version of the CATSv1 model, adds 33 new multiply-imaged galaxies, for a total of 159 individual lensed images.\\
Diego\tablenotemark{m} & 15 &  48 & 4 & 11  & \citealt{Diego:2014b} & {\tt WSLAP+}\tablenotemark{n} : Free-form strong-lensing model using a grid-based method, supplemented by deflections fixed to cluster member galaxies.\\
GLAFIC         & 24 &  67 & 3 & 12 &  \citealt{Ishigaki:2015} & Parametric strong-lensing model using v1.0 of the {\tt GLAFIC} code.\tablenotemark{o} \\
Lam(v1)        & 21 &  65 & 4 & 17 &  \citealt{Lam:2014} & Alternative implementation of the {\tt WSLAP+} model, using a different set of strong-lensing constraints and redshifts.\\
\cutinhead{Unblind models : generated after the SN magnification was known}\\
Bradac(v2)\tablenotemark{e}     & 25 &  72 & 7 & 18 &  \citealt{Wang:2015} & Updated version of the {\tt SWUnited} model with new strong-lensing constraints from HFF imaging and GLASS spectra. Errors derived via bootstrap resampling only strong-lensing constraints.\\
Lam(v2)        & 10 &  32 & 5 & 5  &  \citealt{Lam:2014} & Updated version of the {\tt WSLAP+} model, using more selective strong-lensing constraints and GALFIT\tablenotemark{p} models for galaxy mass.\\
CATS(v2.1)     & 55 & 154 & 8 & 1  &  \citealt{Jauzac:2015b} & Updated version of the CATSv2 model, adopting the spec-z constraints used for Bradac(v2).\\
CATS(v2.2)     & 25 &  72 & 8 & 1  &  \citealt{Jauzac:2015b} & Updated version of the CATSv2 model, adopting the spec-z constraints and multiple-image definitions used for Bradac(v2).
\enddata
\tablenotetext{a}{Number of multiply imaged systems used as strong-lensing constraints.}
\tablenotetext{b}{Total number of multiple images used.}
\tablenotetext{c}{Number of multiply imaged systems with spectroscopic redshifts.}
\tablenotetext{d}{Number of multiply imaged systems with photometric redshifts.}
\tablenotetext{e}{\change{Uses weak lensing constraints from HST- and ground-based imaging, as described in \citet{Merten:2011}.}} 
\tablenotetext{f}{Lens models with the reference code ``HFF'' were produced as part of the 
Hubble Frontier Fields lens modeling
program, using arcs identified in HST archival imaging from
\citealt{Merten:2011}, spectroscopic redshifts from
\citealt{Richard:2014}, and ground-based imaging from
\citealt{Cypriano:2004,Okabe:2008,Okabe:2010a,Okabe:2010b}. Details on
the model construction and an interactive model magnification web interface 
are available at
\url{http://archive.stsci.edu/prepds/frontier/lensmodels/} 
}
\tablenotetext{g}{{\tt SWunited} : Strong and Weak lensing United; \citealt{Bradac:2005}}
\tablenotetext{h}{{\tt CATS} : Clusters As TelescopeS lens modeling team. PI's: J.-P. Kneib \&\ P. Natarajan}
\tablenotetext{i}{{\tt LENSTOOL} : \citealt{Jullo:2007}; \url{http://projects.lam.fr/repos/lenstool/wiki}}
\tablenotetext{j}{\change{Uses weak lensing constraints from HST imaging, as described in \citet{Richard:2014}}}
\tablenotetext{k}{{\tt SaWLENS} : \citealt{Merten:2009}; Strong and Weak LENSing analysis code. \url{http://www.julianmerten.net/codes.html}}
\tablenotetext{l}{{\tt GRALE} : GRAvitational LEnsing; \citealt{Liesenborgs:2006,Liesenborgs:2007,Mohammed:2014}}
\tablenotetext{m}{Abell 2744 model available at \url{http://www.ifca.unican.es/users/jdiego/LensExplorer}. 
No uncertainty estimates were available for the Diego implementation of the {\tt WSLAP+} model, so we adopt the uncertainties from the closely related Lam model.
}
\tablenotetext{n}{{\tt WSLAP+} : \citealt{Sendra:2014}; Weak and Strong Lensing Analysis Package plus member galaxies (Note: no weak-lensing constraints used for Abell 2744)}
\tablenotetext{o}{{\tt GLAFIC} : \citealt{Oguri:2010}; \url{http://www.slac.stanford.edu/~oguri/glafic/}}
\tablenotetext{p}{{\tt GALFIT} : Two-dimensional galaxy fitting algorithm \citep{Peng:2002}}
\end{deluxetable*}

\begin{deluxetable}{p{0.62in}ccc}
\tablecolumns{4}
\tablecaption{Lens model predictions for SN \tomas\ magnification\label{tab:PredictedMagnifications}}
\tablehead{ \colhead{Model\tablenotemark{a}} & 
            \colhead{Best\tablenotemark{b}} & \colhead{Median\tablenotemark{c}} & 
            \colhead{68\% Conf. Range\tablenotemark{d}}}
\startdata
Bradac(v1)      & 3.19     & 2.48  &   2.31$-$2.66 \\
CATS(v1)        & 2.28     & 2.29  &   2.25$-$2.34 \\
CATS(v1.1)      & \nodata  & 2.62  &   2.44$-$2.80 \\
Merten          & 2.33     & 2.24  &   2.04$-$2.92 \\
Sharon(v1)      & 2.56     & 2.60  &   2.44$-$2.78 \\
Sharon(v2)      & 2.74     & 2.59  &   2.42$-$2.85 \\
Zitrin-LTM      & 2.67     & 2.99  &   2.61$-$3.77 \\
Zitrin-NFW      & 2.09     & 2.29  &   2.07$-$2.52 \\
Williams        & 2.70     & 2.81  &   1.65$-$5.54 \\[0.1em]
\tableline\\[-0.5em]
CATS(v2)        & \nodata  & 3.42  &   3.27$-$3.58 \\
Diego           &  \nodata  & 1.80  &   1.44$-$2.16 \\
GLAFIC          & 2.34     & 2.29  &   2.19$-$2.37 \\
Lam(v1)         & \nodata  & 2.79  &   2.42$-$3.16 \\[0.5em]
Bradac(v2)$^{\rm *}$ & 2.23     & 2.26  &   2.30$-$2.23 \\
Lam(v2)$^{\rm *}$ & 1.86     & 1.91  &   1.54$-$2.28\\
CATS(v2.1)$^{\rm *}$ & \nodata & 3.06  &   2.92$-$3.19 \\
CATS(v2.2)$^{\rm *}$ & \nodata & 3.07  &   2.94$-$3.20 \\
\enddata
\tablenotetext{a}{Models above the line are from the pre-HFF set, and those below incorporate HFF data. The final \change{four, marked with asterisks, were not formally} part of the blind test as they include modifications made after the measured magnification of the SN was known.}  
\tablenotetext{b}{The magnification returned for the optimal version of each model, as independently defined by each lens modeling team.}  
\tablenotetext{c}{Median magnification from 100-600 Monte Carlo realizations of the model.}  
\tablenotetext{d}{Confidence ranges about the median, enclosing 68\%\ of the realized values.}
\end{deluxetable}

Table~\ref{tab:PredictedMagnifications} gives each model's predicted
magnification and uncertainty for a source at $z=1.3457$ and at the
position of \tomas.  
For thirteen of these models the SN magnification qualifies as a true
``blind test'', as they were completed before the SN \tomas\
magnification measurement was known.  The final four models
(Bradac-v2, Lam-v2, CATSv2.1 and 2.2) are technically not blind,
although none of these modelers used the SN magnification as a
constraint, and the modelers did not have access to the final SN
magnification value when constructing their model.

These models represent a broad sampling of the
techniques and assumptions that can be applied to the modeling of mass
distributions in galaxy clusters. 
We discuss \change{two primary modeling choices here}, but
a more rigorous comparison of these diverse lens modeling techniques
is beyond the scope of this work. For a complete discussion of each
model's methodology the reader is directed to the listed references.

\smallskip
{\it (1) Parametric vs. Free-form:}
\change{Ten of these models} are so-called {\it parametric}\footnote{Although this
nomenclature is becoming standard in the literature, it is somewhat
misleading, as the pixels or grid cells in free-form models are
effectively parameters as well. Perhaps ``simply-parameterized'' would
be more accurate, though we adopt the more common usage here.}
models, and the remaining seven are {\it free-form} models.  
Broadly speaking, the parametric models use parameterized density
distributions to describe the arrangement of mass within the cluster.
Therefore, parametric models rely (to
varying degrees) on the assumption that the cluster's dark matter can
be described by analytic forms such as NFW halos \citep{Navarro:1997},
or pseudo isothermal elliptical mass
distributions \citep[PIEMD][]{Kassiola:1993}.

The seven free-form models divide the cluster field into a grid,
generally using a multi-scale grid to get better sampling in regions
with a higher density of information (e.g. density of multiple
images). Each grid cell is assigned a mass or a potential, and then
the mass values are iteratively refined to match the observed lensing
constraints. In some cases an adaptive grid is used so that the grid
spacing itself can also be modified as the model is
iterated \citep[e.g.][]{Liesenborgs:2006,Merten:2009,Bradac:2009}.

As usual, there is a tradeoff between a model's flexibility, the
strength of the model assumptions, and the resulting uncertainties. In
a probabilistic framework, the posterior distribution function of the
desired quantities depends on all the priors, including model
assumptions like parametrization.  In general, free-form methods tend
to be more flexibile than simply parameterized models, and thus tend
to result in larger error bars.  In brief, if the free form methods
are too flexible, then they will result in overestimated error
bars. Conversely, if the simply parameterized models are too
inflexible, they will result in underestimated error bars.

\smallskip
\change{{\it (2) Strong vs. Strong+Weak:}}
\change{Thirteen of the mass models evaluated here are constrained only using
strong-lensing features such as multiply-imaged background galaxies
and highly magnified arcs.  There are now $\sim$150 known lensed
images behind Abell 2744 \citep{Jauzac:2015b}, but SN \tomas\ is
located several arcseconds outside the core region of the cluster
where these multiply-imaged galaxies are found.  Therefore these
``strong-lensing only'' models must necessarily rely on extrapolations
to provide a prediction for the SN \tomas\ magnification.}

\change{The other four models in our comparison set also use the strong
lensing constraints from the core region, but additionaly use
weak lensing measurements to provide additional constraints on the
mass distribution farther from the cluster core.  Notes on the
derivation of these weak-lensing constraints are given in the final
column of Table~\ref{tab:LensModels}.  The signal from weak lensing
relies on a large sample of background galaxies, so this constraint
operates principally at separations more than 1\arcmin\ from the
cluster core.  At a projected separation of $\sim40\arcsec$,
SN \tomas\ falls in between the strong- and weak-lensing regimes. }

\begin{figure}
\begin{center}
\includegraphics[width=\columnwidth]{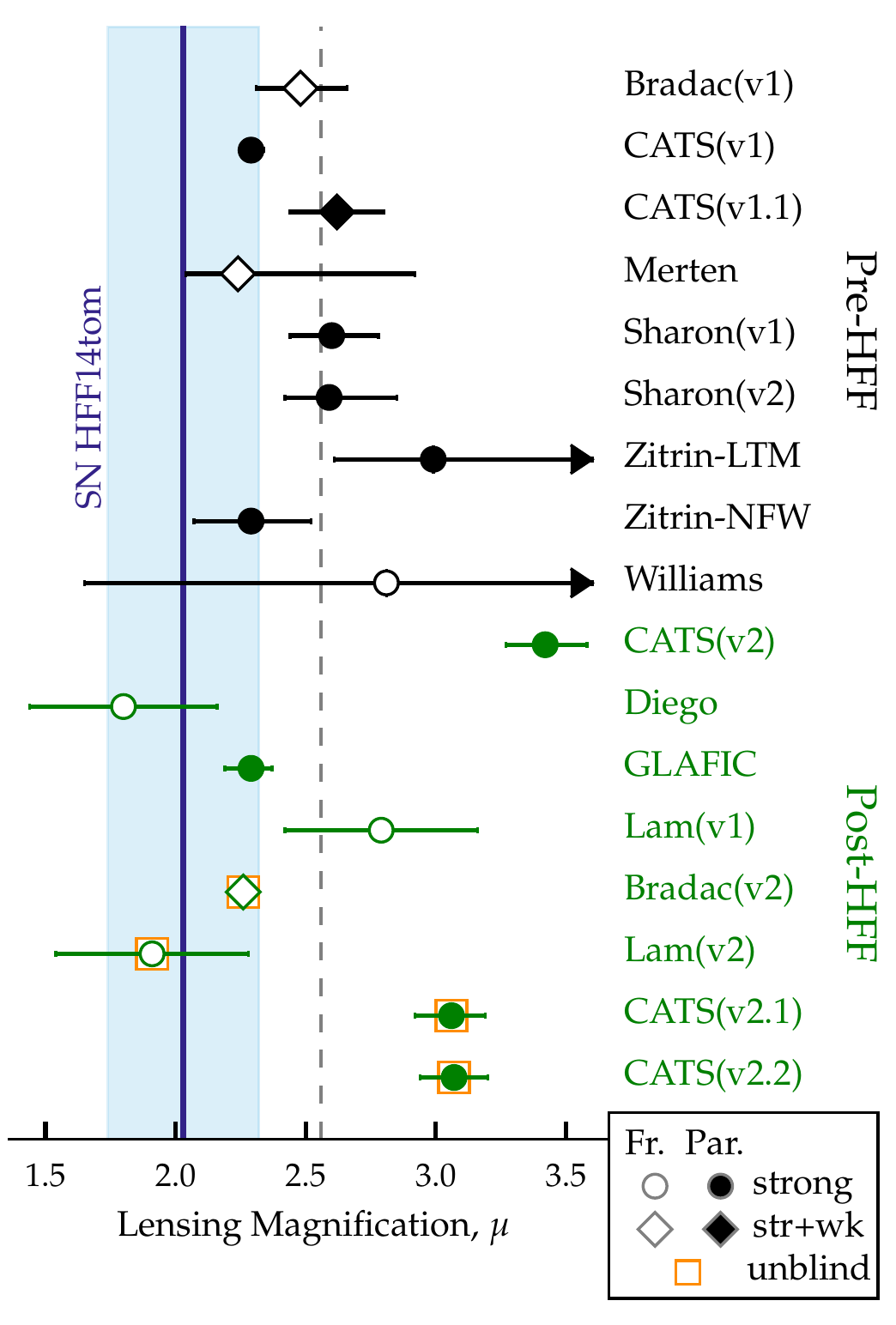}
\caption{ 
Comparison of the observed lensing magnification to predictions from
lens models. The vertical blue line shows the constraints from
SN \tomas\ derived in Section~\ref{sec:DistanceAndMagnification} using
the MLCS2k2 fitter, with a shaded region marking the total
uncertainty. Markers with horizontal error bars show the median
magnification and 68\%\ confidence region from each of the 17 lensing
models.  Circles indicate models that use only strong-lensing
constraints, while diamonds denote those that also incorporate
weak-lensing measurements.  Models using a ``free-form''
approach are shown as open markers, while those in the ``parametric''
family are given filled markers.  The top half, with points in black, shows the nine
models that were constructed using only data available before the
start of the Frontier Fields observations. 
The lower \change{eight} models in
green used additional input constraints, including new multiply imaged
systems and redshifts.  The final four points, with square orange
outlines, are the ``unblind'' models that were generated after the
magnification of the SN was known.  
\change{The black dashed line marks the
unweighted mean for all 17 models, at $\mu=2.6$.}
\label{fig:LensingTest} }
\end{center}
\end{figure}

\medskip
In Figure~\ref{fig:LensingTest} the model predictions are plotted
alongside the observed magnification of SN \tomas, derived in
Section~\ref{sec:DistanceAndMagnification}.  To first order, this
comparison shows that these 17 models are largely consistent with each
other and with the observed magnification of SN \tomas. The
``naive mean'' of the full set of models is $\mu_{\rm pre}=2.6\pm0.4$.
This is an unweighted mean (i.e. we ignore all quoted uncertainties)
derived by naively treating each as an independent
prediction for the magnification (this is clearly incorrect, as
several models are represented multiple times as different versions).
The naive mean is separated from the observed SN magnification by
$\delta\mu/\mu=28\%$, which is approximately a 1.5$\sigma$ difference.

This is approximately consistent with the results
of \P14\ and \citet{Nordin:2014}, where
model predictions were found to be in reasonably good agreement with a
set of 3 lensed SNe from the CLASH program.  The general agreement
between the model predictions and the SN measurement is especially
encouraging for these Abell 2744 models. This is a merging cluster
with a complex mass distribution, and the SN is located
outside of the strong-lensing region where the models are most tightly
constrained.

However, beyond this first-order agreement, there is a small
systematic bias apparent. All but two of the lens models return median
magnifications that are {\it higher} than the observed value, and six
of the models are discrepant by more than $1.5\sigma$. These
six discrepant models are all biased to higher magnifications. They
are found in both the pre-HFF and post-HFF models, in the parametric
and free-form families, and among the strong-lensing-only and the
strong+weak subsets.  It is important to emphasize that SN \tomas\
only samples a single line of sight through the cluster, and this bias
to higher magnifications is minor.  Nevertheless, a systematic shift
of this nature is surprising, given the wide range of modeling
strategies, input data, and physical assumptions represented by this
set of models.  In the following subsections we examine possible
explanations for this small but nearly universal bias.  We first
consider whether a misinterpretation of the data on the SN itself can
account for the observed systematic bias, and then examine the lens
models.

\subsection{Possible Errors in Supernova Analysis}
\label{sec:SupernovaError}

\subsubsection{Redshift Error}
\label{sec:RedshiftError}

If the redshift of the SN derived in Section~\ref{sec:Spectroscopy}
were incorrect, then one would derive a different value for the
magnification, both from the SN measurement and the lens model
predictions.  Conceivably, this could resolve the tension between the
measurement and the models.  It is often the case in SN surveys that
redshifts are assigned based on a host galaxy association, typically
inferred from the projected separation between the SN and nearby
galaxies.  In this case the redshift is strongly supported by
evidence from the SN
itself: we find a consistent redshift from both the SN spectrum
(Section~\ref{sec:SpecRedshift}) and the light curve
Section~\ref{sec:PhotometricClassification}, which are both within
$1\sigma$ of the spectroscopic redshift for the nearest detected
galaxy: $z=1.3457$. This appears to be a solid and self-consistent
picture, so the evidence strongly disfavors any redshift that is
significantly different from $z=1.35$.

We have adopted the most precise redshift of $z=1.3457$ from
the host galaxy as our baseline for the magnification comparison.  If
instead we adopt the spectroscopic redshift from the SN itself
($z=1.31$; Section~\ref{sec:SpecRedshift}) then we find no significant
change in the inferred magnifications or in the suggestion of a small
systematic bias.

\subsubsection{Foreground Dust and SN Color}
\label{sec:ForegroundDust}

All SN sight-lines must intersect some amount of foreground dust from
the immediate circumstellar environment, the host galaxy, and the
intergalactic medium (IGM). In the case of SN \tomas\ one might posit
some dust extinction from the intra-cluster medium (ICM) of Abell
2744, although measurements of rich clusters suggest that the ICM has
only a negligible dust
content \citep{Maoz:1995,Stickel:2002,Bai:2007}.  When fitting
the \tomas\ light curve we account for dust by including corrections
that modify the inferred luminosity distance based on the SN color.
If after applying these dust corrections we were still {\it
underestimating} the effect of dust along this sight-line, then the SN
would appear dimmer than it really is, the inferred distance modulus would be higher,
and the measured magnification would be biased to an artifically low
value. Thus, an underestimation of dust would be in the right
direction to match discrepancy we observe.

In Section~\ref{sec:LightCurveFitting} we found that SN \tomas\ is on
the blue end of the normal range of Type Ia SN colors.  With the SALT2
fitter we measured a color parameter $c=-0.127\pm0.025$, and with
MLCS2k2 we found the host galaxy dust extinction to be $A_{\rm
V}=0.011\pm0.025$ magnitudes.  These colors are tightly constrained,
as we are fitting to photometry that covers a rest-frame wavelength
range from $\sim3500-7000$\AA\ and extends to $\sim$30 days past
maximum brightness.  This leaves little room for the luminosity
measurement to be biased by dust, as the dimming of \tomas\ would also
necessarily be accompanied by some degree of reddening. 

Nevertheless, one might suppose that a bias could be introduced if we
have adopted incorrect values for the color correction parameter
$\beta$ or extinction law $R_V$ in the SALT2 and MLCS2k2 fits,
respectively.  The appropriate value to use for this color correction
and how it affects inferences about the intrinsic scatter in Type Ia
SN luminosities is a complex question that is beyond the scope of this
work \citep[see
e.g.][]{Marriner:2011,Chotard:2011,Kessler:2013,Scolnic:2014a}.
However, we can already rule this out as a solution for the
magnification discrepancy.  Our error on the \tomas\ distance modulus
already includes an uncertainty in the extinction law and a related
error to account for the intrinsic luminosity scatter. These are well
vetted parameters, based on observations of $\sim500$ SNe extending
to $z\sim1.5$ \citep{Sullivan:2011}.  Furthermore, there is no reason
to propose that SN \tomas\ is uniquely affected by a peculiar type of
dust.  Thus, any change in the color correction applied to SN \tomas\
would require the same adjustment to be applied to the unlensed SNe at
similar redshift that make up our comparison sample, largely negating
the effect on the inferred magnification. 

The traditional color corrections as formalized in SN light curve
fitters are designed to account for a dust component that lies in the
rest frame of the SN.  The inferred luminosity of a SN can also be
affected by the presence of foreground dust with a different redshift
and possibly a different reddening law \citep{Menard:2010b}.  However,
the magnitude of such a bias is insufficient to account for the
observed discrepancy, \citet{Menard:2010a} estimate the opacity of the
universe as $\langle A_{V}\rangle\sim0.03$ mag up to $z=0.5$.  While
this can have a measurable impact on precise cosmological constraints,
it is far less than the 0.23 mag difference between the observed
magnification of \tomas\ and the mean of the model predictions.

Although the very blue color of SN \tomas\ is helpful to rule out dust
as a possible explanation for the observed small systematic bias, it
is possible that this very blue color is itself leading to a bias in
the distance measurement.  \citet{Scolnic:2014b} measured a small bias
in the SALT2 fitter for Type Ia SNe that have a color parameter
derived from the light curve fit $c<-0.1$ (see, e.g., their Figure 10
and Section 5.1). For \tomas\ we have measured $c=-0.13$, which would
correspond to a bias of roughly $+0.05$ mag in the SALT2 distance
measurement.  If we were to apply a $-0.05$ mag correction to the
distance modulus from the SALT2 fit, then this would increase the
inferred magnification to $\mu_{\rm SALT2}=2.08\pm0.36$.  This would
slightly reduce the tension between the SALT2 measurement and the model predictions, though it would not be enough to completely alleviate it.

\subsubsection{Misclassification}
\label{sec:Misclassification}

\begin{figure}
\begin{center}
\includegraphics[width=\columnwidth]{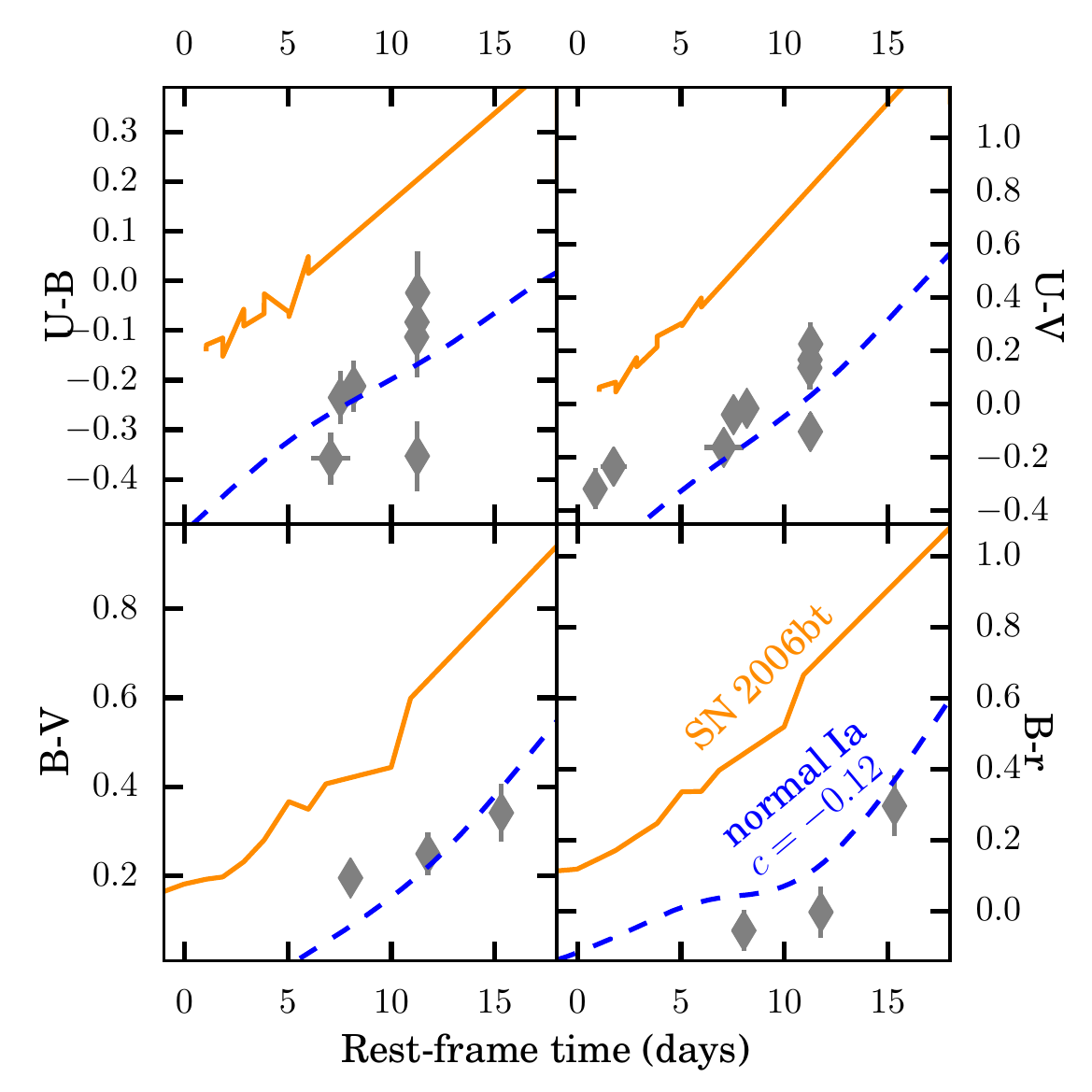}
\caption{ 
Comparing \tomas\ observations to the color curves of SN 2006bt and a
normal Type Ia SN model. Grey points show the observed colors of
SN \tomas, k-corrected to rest-frame UBVr bands. Solid orange lines
show the observed color curves for the peculiar Type Ia SN
2006bt. Dashed blue lines show the color curves for a normal Type Ia
SN, derived from the SALT2 model with a color parameter $c=-0.13$, set
to match the best-fit color for SN \tomas.
\label{fig:colorcomparison} }
\end{center}
\end{figure}

Is it possible that \tomas\ is an example of a peculiar stellar
explosion that does not follow the relationship between light curve
shape and luminosity observed in normal Type Ia SNe?  In
Sections~\ref{sec:SNID} and \ref{sec:PhotometricClassification} we
classified SN \tomas\ as a normal Type Ia SN based on both the
spectroscopic and photometric evidence.  This rules out the
possibility that \tomas\ belongs to a different class of normal SN
explosions (Type Ib, Ic, or II). In
Section \ref{sec:LightCurveFitting} we fit the \tomas\ light curve to
determine a luminosity distance and found that the light curve shape
and color are consistent with a Type Ia SN of average light curve
width, with minimal dust extinction.  This excludes the possibility
that \tomas\ is one of the sub-class of faint and fast-declining Type
Ia SNe like SN 1991bg.

One remaining possibility is that \tomas\ could be part of a rare
sub-category of peculiar Type Ia SNe that masquerade as their normal
cousins, epitomized by the prototype SN 2006bt \citep{Foley:2010}.
These objects appear to have a normal Type Ia light curve shape,
except for the absence of a secondary maximum or ``shoulder'' in
near-IR bands.  The reddest filter available for the \tomas\ light
curve is F160W, which has an effective rest-frame wavelength of
6623\AA\ at $z=1.31$, making it close to the rest-frame {\it r} band.
Although the near-IR shoulder is more prominent in the rest-frame {\it
i} band, we can see in Figure~\ref{fig:LightCurveFits} that the F160W
light curve may suggest a weak or delayed near-IR shoulder.  The
second-to-last observation in F160W is $\sim1\sigma$ lower than
predicted by the best-fit SALT2 and MLCS2k2 models. This is tenuous
evidence, but it would be consistent with a 2006bt-like light curve.

In this case, however, the color of SN \tomas\ can rule out a
classification as a 2006bt-like object.  All members of this peculiar
sub-class exhibit very red colors across all optical and near-IR
bands, consistent with a very cool photosphere \citep{Foley:2010}.
Figure~\ref{fig:colorcomparison} shows that \tomas\ is bluer than the
SN 2006bt prototype by at least 0.25 magnitudes in every color. The
observed \tomas\ colors are fully consistent with a normal, blue
Type Ia SN (the SALT2 model is shown).

\subsection{Possible Errors in Lens Modeling}
\label{sec:LensModelErrors}

Having found no evidence for misinterpetations of the SN
data, we are left to seek an explanation for the mild tension by
scrutinizing the lens models.  Here we evaluate four
possible ways in which the input data or assumptions of these
lens models could lead to a systematic bias in the magnification.

\change{\subsubsection{Unconstrained Mass Model Extrapolation}}

\change{As noted above, the position 
of SN \tomas\ just beyond the edge of the strong-lensing region means
that all models relying only on strong-lensing constraints must
extrapolate the mass profile in order to make a prediction for
the \tomas\ magnification.  Let us suppose that the true cluster
density profile for Abell 2744 happens to have a sharp drop right at
the edge of this core strong-lensing region. In that case these {\it
strong-lensing only} models might well overestimate the mass interior
to the \tomas\ position, and thus systematically overestimate the
magnification.}

\change{ One might expect that models incorporating weak-lensing
constraints would be less susceptible to such bias from a sharply
steepening mass profile.  However, we find that collectively the four
strong+weak lensing models (shown as diamonds in
Figure~\ref{fig:LensingTest}) exhibit the same propensity to
overestimate the magnification along this line of sight.  This
comparison therefore provides no evidence to support the extrapolation
error hypothesis. }

\change{With two versions of the CATS(v1) model, we also have a 
more direct test of the effect of introducing weak lensing
constraints.  The initial CATS(v1) model uses only strong-lensing
constraints, and gives $\mu=2.29^{+0.05}_{-0.04}$, slightly higher
than the measured value.  A second iteration of this model, labeled
here as CATS(v1.1), used the same strong-lensing features, but added
in weak lensing constraints.  The revised magnification of
$\mu=2.62\pm0.18$ is further from the measured value, which serves to
reinforce the conclusion that adding weak lensing constraints cannot,
by itself, resolve this bias.}

\change{Given the intermediate location of this SN, caught between the
strong-lensing and weak-lensing regimes, an even more useful tool than
weak lensing would be the {\it flexion}
signal \citep{Goldberg:2002,Goldberg:2005,Irwin:2006,Goldberg:2007}.
Flexion is a second-order lensing effect that distorts the shapes of
background galaxies that are not lensed strongly enough to be warped
into arcs or filaments.  A flexion measurement near the SN \tomas\
sight line would be more sensitive to local gradients and substructure
in the cluster mass profile.  However, no such flexion constraints are
yet available.}

\change{\subsubsection{Line of Sight Structure}}
In addition to the dominant gravitational lensing from the foreground
cluster, a distant source like SN \tomas\ will also be subject to the
cumulative weak lensing effects due to uncorrelated large scale
structure (LSS) along the line of sight, sometimes called ``Cosmic
Weak
Lensing'' \citep{Wong:2011,Host:2012,Collett:2013,Greene:2013,Bayliss:2014,McCully:2014,DAloisio:2014}.
This \change{LSS} effect is most important for very high redshift
sources ($z>5$), which have a much longer path length over which to
encounter large scale structure and for highly magnified sources very
near to the cluster lensing critical curve (because \change{LSS} can
perturb the position of the critical curve).  For SN \tomas\ at the
modest redshift of $z=1.35$ and far from the critical
curve, \change{LSS} should increase the scatter in the lensing
magnification at a level much less than 10\% \citep{DAloisio:2014}.
The \change{LSS} effect is therefore unable to account for the
observed discrepancy of $\sim23$\% by itself.

In addition, some sources of uncertainty are taken into account only by a
subset of models. For example, the Willams model (using GRALE;
Liesenborgs et al. 2006, 2007; Mohammed et al. 2014) accounts for mass
sheet degeneracy by introducing a free parameter in the reconstruction
that represents an arbitrarily scaled mass sheet. The flexibility of
the Williams model and the inclusion of this source of uncertainty
contribute to its relatively large uncertainties compared to the other
models.

\subsubsection{Nearby Cluster Member}
A component of the Abell 2744 lens that is particularly relevant to SN \tomas\ is the cluster member
galaxy that lies just 5\farcs8 north of the SN position (see
Figure~\ref{fig:DiscoveryImage}).  If the mass-to-light ratio or the
profile of the dark matter halo for this galaxy were significantly
different from other cluster member galaxies, then its
proximity to the SN sight-line might drive a bias in the magnification.
We tested this hypothesis using the CATS(v2) model by allowing the mass
of the nearby cluster member galaxy to vary as a free parameter in the
model. We found that the change in the SN \tomas\ magnification
prediction was less than $\Delta\mu=0.1$.  This additional dispersion
is already included in the uncertainties quoted for that model in
Table~\ref{tab:PredictedMagnifications} and
Figure~\ref{fig:LensingTest}.  Furthermore, an erroneous M/L
value for the nearby cluster member galaxy could not explain the
systematic shift of all lens models, because some do not incorporate
cluster member galaxies into their constraints at all (the free-form
Bradac, Williams and Merten models).

\subsubsection{Cosmological Parameter Uncertainty}
Most lens models make a
fixed assumption for the values of cosmological parameters in a
standard \LCDM\ cosmology (typically: \Om=0.3,\OL=0.7,\Ho=70 km
s$^{-1}$ Mpc$^{-1}$). This can introduce a systematic magnification
error that is comparable in magnitude to that of the statistical
uncertainties \citep{Zitrin:2014b,Bayliss:2015}.  
Incorporating this
cosmological parameter uncertainty would increase the model magnification 
errors, and therefore reduce the tension between models and observation, but would not
resolve the overall systematic shift.  \citet{Bayliss:2015} find that for Abell 2744 the additional cosmological uncertainty for magnifications $\mu\approx2$ is $<10\%$ which would still leave 6 models discrepant by at least 1.5$\sigma$.

\subsubsection{Misidentification of a multiple image}
\citet{Jauzac:2015b} proposed a correction
for the location of image 3.3, the third component of a triply-imaged
galaxy \citep{Merten:2011} at $z=3.98\pm0.02$ \citep{Johnson:2014}.
This possibly specious multiple image is only $\sim$10\arcsec\ from
the SN \tomas\ position, meaning a change in its location could have a
large impact on the predicted magnification for
SN \tomas\ \citep[see][for a quantitative discussion of this
effect]{Johnson:2014}.  However, \citet{Jauzac:2015b} also evaluated a
version of the CATS(v2) model in which this single source is left in
the original position and found that the location of image 3.3 does
not significantly affect the predicted magnifications.  Furthermore,
dividing our 17 models into those using the original location vs those
that adopt the new position, we find that both groups include models
that are within 1$\sigma$ of the observed $\mu$ as well as
significantly discrepant points.

\subsubsection{Source Plane Minimization}
\label{sec:SourcePlaneMinimization}

A general tendency towards high model magnification can result when
lens models are minimized in the source plane (rather than the image
plane).  For a given model, modifying the surface mass density to have
a shallower profile (declining less quickly with radius) will lead to
an increase in the predicted magnifications across the field.  As the
magnification applies to both flux and area, the shallower profile
also leads to a smaller source plane area, which necessarily brings
delensed images of the same source closer together in the source
plane.  Thus, models that are \change{naively} optimized in the source
plane will have a tendency toward shallower mass profiles and larger
magnifications, which force all sources toward the same location in
the source plane 
\change{\citep[see, e.g.,][]{Kochanek:1991a,Kochanek:2006,Jee:2007,Ponente:2011}.}

There are several strategies for mitigating this well known bias.
Optimizing a model in the image plane can avoid this bias
entirely \citep{Broadhurst:2005,Zitrin:2009a}. The Sharon(v2),
Zitrin-LTM, Zitrin-NFW, and CATS(v1,v2-2.2) models all use image plane
minimization.  Alternatively, one may guard against this bias by using
a modified source plane minimization that is constrained not to
generate overly small de-lensed
sources \citep{Sendra:2014,Diego:2014}. In our test set, a version of
this second strategy is employed by the \change{GLAFIC, Williams, and
Diego} models.  \change{For example, the Williams model uses source
plane optimization in the GRALE algorithm, but avoids bias by
maximizing the {\it fractional} overlap of extended source plane
images.}  Finally, the effect of this bias should be reduced when one
uses many strong lensing constraints over a wide range of
redshifts. The CATS(v2) and CATS(v2.1) models have the largest number
of strong-lensing constraints, using 50 and 55 multiply-imaged
systems, respectively.

Although most of the tested models utilize one of these 
strategies to prevent a source plane minimization bias,
Table~\ref{tab:PredictedMagnifications} and
Figure~\ref{fig:LensingTest} show that these models do not uniformly
deliver more accurate magnification predictions.  The most direct test
of this is in the Sharon and CATS model series.  
\change{Both of these model series use the LENSTOOL 
software \citep{Kneib:1996,Jullo:2007,Jullo:2009,Kneib:2011a}, which
allows a choice of how the minimization is done.  For the Sharon
series, the v1 model used the faster source plane minimization
approach, but the v2 model was constructed with the more accurate
image plane minimization.  The CATS v1, 2, 2.1, and 2.2 models all
used image plane optimization, while the v1.1 employed source plane
optimization (and also added in weak lensing constraints).  In neither
of these model series do we see a uniform improvement in accuracy when
optimization is done in the image plane.  This test strongly suggests
that the observed tension between the lens models and the SN
observation cannot be entirely attributed to a bias arising from
source plane minimization.}

\subsection{Quantity and Quality of Strong Lensing Constraints}
\label{sec:QuantityAndQuality}

None of the possible errors in lens model inputs described above can
completely account for the mild tension between the observed and
predicted magnifications.  However, we do have several models that
agree with the measured SN magnification.  In this section we take a
simple pragmatic approach and seek to identify any shared
characteristics of the models that are most accurate for this
particular sight-line.  In this way SN \tomas\ may serve as a guide
toward optimizing future lens models for magnification predictions,
particularly in regions where strong-lensing constratins may be
scarce.

\begin{figure*}
\begin{center}
\includegraphics[width=\textwidth]{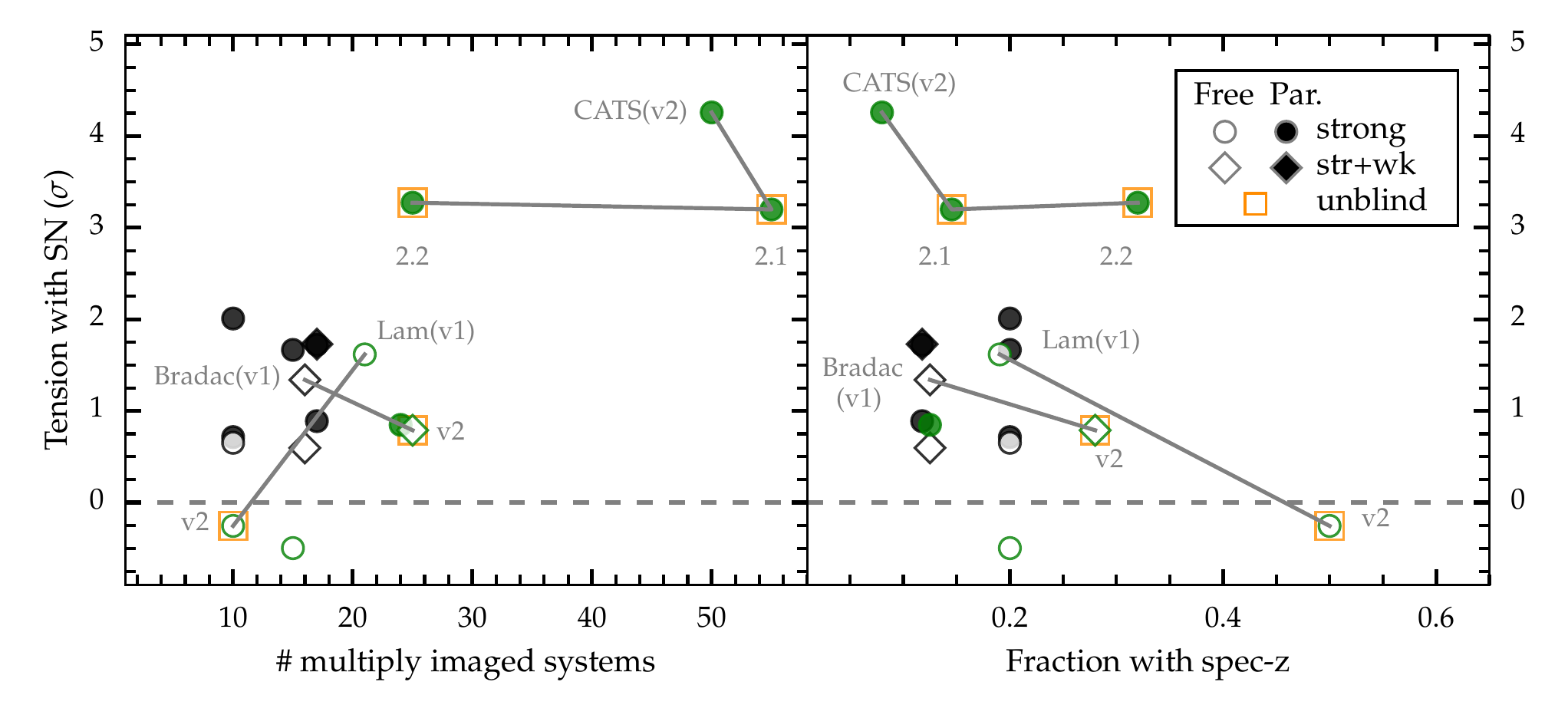}
\caption{ \label{fig:Nspecz}
The model tension as a function of strong-lensing constraints. The y
axis in both panels marks the number of standard deviations in
discrepancy between the measured SN magnification and the prediction
from each lens model\change{: $(\mu_{\rm model}-\mu_{\rm SN})/(\sigma_{\rm model}^2 + \sigma_{\rm SN}^2)^{1/2}$.}  The left panel plots this tension against the
total number of multiply imaged systems used, and the right panel
plots the fraction of
multiply-imaged systems that have a spectroscopic redshift
constraint.  Marker shapes and colors are the same as in
Figure~\ref{fig:LensingTest}.  In both panels, three lens model sequences are highlighted with connecting lines: the Bradac, Lam, and CATS(v2) models, which are the three examples where the number of strong lensing constraints and the fraction with spectroscopic redshifts changed substantially from one model version to the next.}
\end{center}
\end{figure*}

The addition of deeper imaging and new spectroscopic constraints from
the HFF program and affiliated efforts might be expected to improve
the accuracy of magnification predictions. However, in
Figure~\ref{fig:LensingTest} we saw no evidence for a universal
improvement in model accuracy when moving from the pre- to the
post-HFF models.  In Figure~\ref{fig:Nspecz} we attempt to refine this
hypothesis by isolating the quantity and quality of strong lensing constraints. 

The left panel plots the tension between each model prediction and the
measured SN magnification against the number of multiply-imaged
galaxies used as strong-lensing constraints. The Bradac, Lam and
CATSv2 model series are highlighted here with connecting lines, as
three examples where the number of systems used has changed
substantially from one version to the next, but the model construction
has remained fundamentally unchanged.
There is no clear correlation here, and in
fact for both the Lam and CATSv2 model series, we see the tension
actually {\it increases} for a larger number of strong-lensing constraints.

In the right panel of Figure~\ref{fig:Nspecz} we examine an
alternative explanation by plotting the model tension against the
fraction of strong-lensing constraints that have a spectroscopic
redshift. 
For the SN \tomas\ magnification prediction, we see that all three of
the testable model series (Bradac,Lam,CATSv2) become {\it more}
accurate as the spec-z fraction increases.\footnote{Of course this
apparent correlation does not imply any direct causation. Consider the
limit of just a single multiply-imaged galaxy constraint: the spec-$z$
fraction could be 1.0 but the model would be wildly unconstrained.}
The spec-$z$ fraction here serves as a crude stand-in for measuring
how stringently the strong-lensing constraints are selected in these
models.

The SN \tomas\ comparison therefore supports the notion that the {\it
quality} of strong-lensing constraints is just as important as the
{\it quantity} -- at least for computing magnifications of sources
outside the strong-lensing region. Once a sufficient number of robust
multiply-imaged systems have been included (perhaps $\sim$10) most
models will be dominated by systematic uncertainties such as those
described in Section~\ref{sec:LensModelErrors}.  To improve the
accuracy of magnification predictions with a large number of
strong-lensing constraints, one should employ a strict vetting of the
multiply-imaged systems.  Each system used should ideally have a
spectroscopic redshift, or at least a very well constrained
photometric redshift.  The primary value in this strict vetting is to
avoid erroneous redshifts being assigned to multiply-imaged systems
that are unconstrained in redshift space.  Such redshift errors are
especially problematic for relatively low-$z$ background systems: at
lower redshifts the model-derived deflection angle and magnification
show a more rapid variation with the redshift of the background galaxy
than for similarly positioned high redshift galaxies. This conclusion
should be unsurprising, as for example, \citet{Bayliss:2015} have
shown that cosmological uncertainties are smaller for models that use
a large fraction of strong-lensing constraints with spectroscopic
redshifts, and \citet{Johnson:2014} have shown a case study where
adding spectroscopic redshift information changes the inferred
magnification by $\sim$10\%.

It is relevant to note that the three model series highlighted here
(Bradac, CATSv2 and Lam) also happen to be the three models that are
not truly ``blind,'' in that the modelers were aware of preliminary
analysis of the SN magnification when constructing the latest version.
However, none of these models used the SN magnification as an input
constraint, and none were consciously tuned to match the SN
magnification. In fact, for the Bradac(v2) and Lam(v2) cases, the
modelers did not have access to a final measurement of the SN
magnification at the time their models were generated.

\section{Summary and Conclusions}
\label{sec:SummaryAndConclusions}

The appearance of a Type Ia SN behind a massive galaxy cluster
provides a rare opportunity to use a standard candle for a direct
measurement of the absolute magnification due to gravitational
lensing.  The discovery of SN \tomas\ in the HFF imaging of Abell 2744
offers the first chance to apply this test on a cluster with multiple
publicly-available lens models.  We have found that the spectrum and
light curve of SN \tomas\ are well matched by templates of a normal
Type Ia SN at $z=1.3457$.  Using the two most prevalent \SNIa\ light
curve fitters, SALT2 and MLCS2k2, we get a consistent measurement of
the distance modulus (Table~\ref{tab:MeasuredMagnification}).  Using a
cosmology-independent comparison against a sample of unlensed \SNeIa\
at similar redshifts, we find that SN \tomas\ is $\sim0.7$ magnitudes
brighter than the field sample would predict.  Attributing this
difference to the gravitational lensing magnification (and accounting
for the intrinsic scatter in luminosity of the \SNIa\ population), we
have derived a consistent measured magnification of $\mu_{\rm
SALT2}=1.99\pm0.38$.  $\mu_{\rm MLCS2k2}=2.03\pm0.29$, from the two
light curve fitters.

Taking advantage of the availability of 17 well-constrained lens
models for the Abell 2744 cluster, we have used SN \tomas\ to ask how
accurately these lens models can predict the magnification along this
line of sight.  We find that these models are consistent, and fairly
accurate, collectively predicting $\mu=2.5\pm0.4$, within 1$\sigma$ of
the measured value.  This is encouraging, and reinforces the quality
and value of these public lens models for studying magnified
background objects.  However, we note that all but two of the tested
models provide a median magnification prediction that is larger than
the measurement from the SN, and all the models with significant
discrepancy are too high.  This may be an indication that there is a
small systematic bias at work, at least in some of the models. 

We have speculated on what could be the origin for such a systematic
bias, first considering and rejecting three possible explanations that
presume an error in the interpretation of the available SN data.

\begin{enumerate}
\item Redshift: We reject the possibility that a redshift error is the primary cause
of the discrepancy, as the redshift evidence is well supported by
multiple lines of evidence from both the SN and the presumed host
galaxy. Changing the redshift within the constraints of these
complementary observations does not resolve the tension. 
\item Dust: We find
it implausible that there is sufficient dust in the cluster or
elsewhere along the line of sight to account for the magnification
discrepancy.  A dust-induced bias is also disfavored by the very blue
color of SN \tomas.
\item Misclassification: The combination of spectroscopic and
photometric evidence strongly supports our classification of
SN \tomas\ as a normal Type Ia SN.  The most plausible
mis-classification would be that the object is a peculiar Type Ia of
the SN 2006bt-like sub-class.  Although the light curve shape could
allow this possibility, the blue color of \tomas\ can once again
reject this alternative. 
\end{enumerate}

\noindent 

Turning to the lens models, we have considered 6 ways in which a bias
might be introduced into the lens models:

\begin{enumerate}
\item \change{Mass profile extrapolation errors}
\item \change{Large scale structure along the line of sight}
\item Peculiar mass for a nearby cluster member
\item Cosmological parameter uncertainty
\item Misidentification of a multiple image
\item Source plane minimization
\end{enumerate}

\noindent
Although each of these could plausibly introduce a small systematic
bias, we find that none are likely to be significant enough (or
universal enough) to completely resolve the tension between models and
observations.  

Finally, we have examined whether there is a simple prescription for
the kind of strong lensing constraints that are most likely to deliver
an accurate magnification prediction for this sight-line.  We find
that the number of multiply imaged systems used is not in itself
predictive of model accuracy.  Rather, it is the fraction of multiply
imaged systems that have spectroscopic redshifts that is most
correlated with model accuracy.  A reasonable interpretation -- at
least for this cluster and this particular set of models -- is that
the spectroscopic redshift fraction serves as an effective proxy for
the ``quality'' of the strong lensing constraints. This quality of the
input data appears to be a key ingredient for deriving accurate
magnifications outside the strong-lensing regime.

We have evaluated here only a single object behind a single cluster,
with a minor tension between the observed magnification and the model
predictions.  This is not in and of itself a cause for alarm.
Previous analyses of lensed \SNIa\ found no significant discrepancy
between the observed \SNIa\ magnifications and the predictions from
lens models (\P14; \citealt{Nordin:2014}), albeit
with a much smaller set of lens models being tested.  The observed
systematic bias for \tomas\ is small, and many of the lens models that
deviate from the measured magnification are preliminary models that
have not been updated to include all of the HFF data.  Future
revisions of the lens models for Abell 2744 could either incorporate
the observed magnification of \tomas\ as a new model constraint, or
can revisit this test to evaluate whether the bias persists.

The general problem of combining predictions from many independent
models has a rich history in the astronomical and statistical
literature \citep[e.g.,][]{Press:1997,Hoeting:1999,Liddle:2009}.
In a recent example, \citet{Dahlen:2013} examined the photometric
redshifts from multiple galaxy SED fitting codes using the same input
data.  They found that the photo-z accuracy and precision can be
substantially improved by combining the outputs using a sum of
probability distributions, a hierarchical Bayesian method, or even a
technique as simple as taking the straight median from all available
models.  It may be that lensing magnifications are similar, in that
they can in general be improved through a simple combination of lens
model predictions.  Such model averaging will not, however, resolve a
{\it systematic} bias that is shared by many models.

\subsection{Future Work}

A promising avenue for exploring the origins of such systematic biases in
cluster lens models is through the use of simulated lensing data.  One
can start with very deep high-resolution multi-band imaging on an
unlensed field that has a fairly complete spectroscopic redshift
catalog, such as the Hubble Ultra Deep Field.  Then a simulated galaxy
cluster is placed in the field, and the background galaxies are
distorted into arcs and multiple images using a well-defined lensing
prescription.  The artificially lensed images can then be distributed
to lens modeling teams who attempt to reconstruct the (known) mass
profile of the simulated cluster.  This exercise has recently been
pursued with a set of synthetic clusters similar to those observed in
the HFF program (Meneghetti et al. in prep), using techniques
similar to \citet{Meneghetti:2010,Meneghetti:2014}.  Preliminary
analysis of this simulation comparison suggests that
magnifications can be systematically overestimated for
sources that lie outside the strong-lensing region along the minor
axis direction of a simulated cluster that is very
elongated, similar to Abell 2744.  This indicates that some
lens models significantly underestimate the ellipticity of the mass
distribution on large scales for that simulated cluster, which is
consistent with a recent lens model comparison using 25 real clusters
from the CLASH program \citep{Zitrin:2015}. Our analysis of
SN \tomas\ here indicates that such simulation efforts and
head-to-head lens model comparisons will be an important step for
moving toward precision science with cluster-lensed sources.  

Increasing the sample of SNe behind clusters like those in the HFF
program -- with rich lensing constraints and deep imaging -- would
allow the test described here to be repeated and refined.  With 10 or
100 such objects, it would be possible to see whether the SN \tomas\
$\mu$ discrepancy is simply an outlier, or an indication of a more
pernicious systematic error.  Any cluster that has been vetted by
pencil-beam magnification tests using lensed \SNIa\ will be able to
provide a more reliable measure of the magnifications for very high
redshift objects.  Similarly, a broad sample of SNe like \tomas\ would
help to define the preferred lens modeling methodology by highlighting
any models that consistently perform well in \SNIa\ lensing tests.
The ongoing FrontierSN program will discover and follow any more
highly magnified SNe that appear behind the Frontier Field
clusters. Unfortunately, the HFF survey is not designed with high-z
transient discovery as a primary science goal, so the FrontierSN
effort will likely add no more than 1-3 new lensed \SNIa. Further
imaging of strong-lensing clusters with \HST\ or the James Webb Space
Telescope (JWST) could enable a larger sample to be collected,
especially if the filters and cadence are optimized for detection
of \SNIa\ at $z>1$.  Massive clusters such as Abell 2744 will continue
to be attractive as cosmic telescopes, allowing the next generation of
telescopes to reach the faintest objects in the very early universe.
The puzzling bias revealed by SN \tomas\ supports a concerted effort
to improve these lenses with further examination of lensing
systematics through simulations and collection of a larger sample of
magnified SNe.

\bigskip

{\bf Acknowledgments:}

This work is dedicated to our colleague and friend Tomas Dahlen, for
whom this supernova has been named in memoriam. He is dearly missed.

We thank the Hubble Frontier Fields team at STScI for their
substantial efforts to make the HFF program successful.  In
particular, thanks are due to Matt Mountain for the allocation of
discretionary orbits for the HFF program; to Jennifer Lotz, Norman
Grogin and Patricia Royle for accommodations in strategy and
implementation to make the FrontierSN program possible; to Anton
Koekemoer for HFF data processing support.  We also thank
the CLASH team, led by Marc Postman, for observations, catalogs, and
high level science products that were of significant value for this
analysis.  Thanks to Jonatan Selsing for helpful comments on the
manuscript.

This work utilizes gravitational lensing models produced by modeling
teams that were funded as part of the HST Frontier Fields program
conducted by STScI. STScI is operated by the Association of
Universities for Research in Astronomy, Inc. under NASA contract NAS
5-26555. The lens models were obtained from the Mikulski Archive for
Space Telescopes (MAST).

Financial support for this work was provided to S.A.R. by NASA through
grants HST-HF-51312 and HST-GO-13386 from STScI, which is operated by
Associated Universities for Research in Astronomy, Inc. (AURA), under
NASA contract NAS 5-26555. A.M. acknowledge the financial support of
the Brazilian funding agency FAPESP (Post-doc fellowship - process
number 2014/11806-9). Support for this research at Rutgers University
was provided in part by NSF CAREER award AST-0847157 to SWJ.  The Dark
Cosmology Centre is supported by the Danish National Research
Foundation. J.M.D acknowledges support of the consolider project
CSD2010-00064 and AYA2012-39475-C02-01 funded by the Ministerio de
Economia y Competitividad. J.M. contributed to this research from the
Jet Propulsion Laboratory, California Institute of Technology, under a
contract with NASA and acknowledges support from NASA Grants
HST-GO-13343.05-A and HST-GO-13386.13-A. The research leading to these
results has received funding from the People Programme (Marie Curie
Actions) of the European Union's Seventh Framework Programme
(FP7/2007-­2013) under REA grant agreement number
627288. A.Z. acknowledges financial support from NASA through grant
HST-HF-51334.01-A awarded by STScI and operated by AURA. 
TT acknowledges support by the Packard Foundation in the form of Packard Research Fellowship. GLASS is funded by NASA through HST grant GO-13459. LLRW
acknowledges the support of the Minnesota Supercomputing Institute.

{\it Facilities:} \facility{HST (WFC3)}
\pagebreak

\end{document}